\documentclass[a4paper,11pt]{article}
\pdfoutput=1
\usepackage{jcappub}
\usepackage{bbold}

\usepackage[english]{babel}
\usepackage[utf8]{inputenc}
\usepackage{amsmath}
\usepackage{color}
\usepackage{amsfonts}
\usepackage{graphicx}
\usepackage{amssymb}
\usepackage{mathtools}
\usepackage{placeins}
\usepackage{float}
\graphicspath{{plots/}}

\def\be{\begin{equation}}
\def\ee{\end{equation}}
\def\bea{\begin{eqnarray}}
\def\eea{\end{eqnarray}}
\def\bean{\begin{eqnarray*}}
\def\eean{\end{eqnarray*}}

\def \De {\Delta}

\def \kvec {\mathbf{k}}
\def \rvec {\mathbf{r}}

\def \khat {\hat{\mathbf{k}} }

\newcommand{\bk}{\boldsymbol{k}}
\newcommand{\bq}{\boldsymbol{q}}
\newcommand{\bn}{\boldsymbol{n}}

\newcommand{\bx}{\boldsymbol{x}}

\newcommand{\bfk}{\mbox{\boldmath$k$}}
\newcommand{\bfp}{\mbox{\boldmath$p$}}
\newcommand{\bfq}{\mbox{\boldmath$q$}}

\newcommand{\bell}{\boldsymbol{\ell}}

\newcommand{\qvec}{\textbf{q}}

\title{Non-linear contributions to angular power spectra}

\author{Mona Jalilvand,}
\author{Basundhara Ghosh,}
\author{Elisabetta Majerotto,}
\author{Benjamin Bose,}
\author{Ruth Durrer,}
\author{Martin Kunz}

\affiliation{
Universit\'e de Gen\`eve, D\'epartement de Physique Th\'eorique and Centre for Astroparticle Physics,
24 quai Ernest-Ansermet, CH-1211 Gen\`eve 4, Switzerland
}

\emailAdd{mona.jalilvand@unige.ch}
\emailAdd{basundhara.ghosh@unige.ch}
\emailAdd{elisabetta.majerotto@unige.ch}
\emailAdd{benjamin.bose@unige.ch}
\emailAdd{ruth.durrer@unige.ch}
\emailAdd{martin.kunz@unige.ch}

\abstract{
 Future galaxy clustering surveys will probe small scales where non-linearities become important. Since the number of modes accessible on intermediate to small scales is very high, having a  precise model at these scales is important especially in the context of discriminating alternative cosmological models from the standard one. In the mildly non-linear regime, such models typically differ from each other, and galaxy clustering data will become very precise on these scales in the near future. As the observable quantity is the angular power spectrum in redshift space, it is important to study the effects of non-linear density and redshift space distortion (RSD) in the angular power spectrum. We compute non-linear contributions to the angular power spectrum using a flat-sky approximation that we introduce in this work, and compare the results  of different perturbative approaches with $N$-body simulations.
 We find that the TNS perturbative approach is significantly closer to the $N$-body result than Eulerian or Lagrangian 1-loop approximations, effective field theory of large scale structure or a halofit-inspired model. However, none of these prescriptions is accurate enough to model the angular power spectrum well into the non-linear regime. In addition, for narrow redshift bins, $\Delta z \lesssim 0.01$, the angular power spectrum acquires non-linear contributions on all scales, right down to $\ell=2$, and is hence not a reliable tool at this time. To overcome this problem, we need to model non-linear RSD terms, for example as TNS does, but for a matter power spectrum that remains reasonably accurate well into the deeply non-linear regime, such as halofit.
}
\begin{document}

\maketitle

\section{Introduction}
\label{Sec1}
\setcounter{equation}{0}
After the tremendous success of Cosmic Microwave Background (CMB) observations \cite{Planck:2015xua}, presently major efforts in  cosmology are going into the observation and modelling of the distribution of galaxies \cite{Anderson:2012sa,Song:2015oza,Beutler:2016arn,Blake:2011rj,Reid:2012sw,Macaulay:2013swa,Beutler:2013yhm,Gil-Marin:2015sqa,Simpson:2015yfa}. As this data set is  three dimensional, it is potentially much richer and may allow us to study the evolution of cosmic structure formation. 

However, on small scales the fluctuations in the matter density can become large at the present time. Therefore, first order cosmological perturbation theory is not sufficient to describe structure formation on these scales and numerical N-body simulations, in principle including also hydrodynamic effects, are needed. This is a very complicated process and usually many phenomenological parameters have to be used to describe the highly non-linear hydrodynamic processes which are affected by star formation, AGN feedback and more~\cite{Rabold:2017qdp,Schneider:2018pfw,Ocvirk:2018pqh}.

On intermediate scales, higher order perturbation theory and  phenomenological modelling of the galaxy power spectrum can be used \cite{Kaiser:1987,Scoccimarro:2004tg,Taruya:2010mx,Okumura:2015fga,Bose:2016qun}. This is the topic of the present work. In the past, people have mainly looked at the power spectrum in Fourier space \cite{Beutler:2013yhm,Beutler:2016arn}. Within linear perturbation theory this is approximated by the so-called Kaiser formula~\cite{Kaiser:1987}, which includes redshift space distortions (RSD) i.e. the fact that the observed redshift is affected by peculiar velocities which are in turn correlated with matter overdensities, 
\bea\label{e:poldk}
P(k, \mu, \bar{z}) &=& D_1^2(\bar{z})\left[b(\bar z) +f(\bar z)\mu^2\right]^2P_m(k), 
\eea
where $\mu = \hat\bk\cdot\bn$ is the cosine of the angle between the unit vector in direction $\bk$, $\hat\bk$, and the observation direction $\bn$, which is a unit vector. Here $\bar z$ is a mean redshift of the survey under consideration, $P_m(k)$ is the linear matter density power spectrum today,  $D_1(\bar z)$ is the linear growth factor normalized to $D_1(0)=1$, $b(\bar z)$ is the galaxy bias and
\be\label{e:growth}
f(\bar z) = -\frac{D_1'}{D_1}(1+\bar z)=\frac{d\ln D_1}{d\ln(a)} \,,
\ee
is the growth rate, where the prime denotes the derivative with respect to the redshift $\bar z$.
This formula has been generalized in the literature to include non-linearities in the matter power spectrum, usually by replacing $D_1^2(\bar z)P_m(k)$  by a 1-loop or 2-loop power spectrum \cite{Scoccimarro:2004tg} or by a phenomenological approximation like halofit~\cite{Takahashi:2012em,Taruya:2010mx}. Workers in the field have also corrected the `Kaiser relation' $f(\bar z)(\hat\bk\cdot \bn)^2$ for the peculiar velocity with a non-linear and phenomenological description~\cite{Markovic:2019sva}.
With the increasing precision of the data available from galaxy surveys such as Euclid\footnote{\url{www.euclid-ec.org}} \cite{Laureijs:2011gra,Amendola:2016saw}, WFIRST \footnote{\url{https://wfirst.gsfc.nasa.gov/}} \cite{Spergel:2015sza}, 4MOST \cite{4MOST:2019} and (DESI)\footnote{\url{www.desi.lbl.gov}}\cite{Aghamousa:2016zmz}, and with the upcoming HI surveys (e.g.\ \cite{Bandura:2014gwa,Newburgh:2016mwi,Bacon:2018dui}) that have a very high redshift resolution,
it is important to model the theoretical galaxy power spectrum as accurately as possible. Even at scales as large as those of baryon acoustic oscillations, we need to go beyond linear perturbation theory~\cite{Beutler:2013yhm,Bose:2019psj}.

Eq.~\eqref{e:poldk} is a good approximation to cosmological observations only if we have a small, far away galaxy survey in a fixed direction $\bn$ at nearly fixed redshift $\bar z$. A true galaxy survey lives on our background lightcone and the radial distance between galaxies is related to their redshift difference. The correlation function therefore is truly a function of two directions, $\bn_1,~\bn_2$ and two redshifts, $z_1,~z_2$. Assuming statistical isotropy it depends only on $\cos\theta=\bn_1\cdot\bn_2$, $z_1$ and $z_2$. A harmonic transform in $\cos\theta$ yields the spherical power spectrum $C_\ell(z_1,z_2$). This has been derived at first order in perturbation theory in~\cite{Bonvin:2011bg,Challinor:2011bk}. Apart from density and RSD, the complete formula includes several relativistic effects like the integrated Sachs Wolfe effect, the Shapiro time delay, the gravitational potential at the source and gravitational lensing convergence (also termed `magnification bias'). Apart from the last term, all relativistic contributions are relevant only on very large scales corresponding to $\ell\lesssim 10$. The gravitational lensing contribution is relevant in wide redshift bins, at relatively high redshifts, $z\gtrsim 1$, or in widely separated redshift bins~\cite{Bonvin:2011bg,Montanari:2015rga,Cardona:2016qxn}. For the redshift bin widths used in this work, we discuss briefly in Appendix \ref{app:lensing} the importance of lensing in angular power spectra, relative to the RSD contribution.

Here, we consider spectroscopic surveys which have a very precise redshift distribution and we shall neglect lensing. We want to determine the effect of loop corrections in Eulerian and Lagrangian perturbation theory as well as other phenomenological approaches to the non-linear matter power spectrum. We study how these corrections affect the observable angular power spectrum, $C_\ell$, when considering  density and redshift space distortions, and we compare them with results from the more accurate COmoving Lagrangian Acceleration (COLA)\cite{Tassev:2013pn,Howlett:2015hfa, Valogiannis:2016ane,Winther:2017jof} simulations. The density and RSD contributions are dominant at relatively low redshifts and for spectroscopic surveys like Euclid; furthermore, it is these terms which are most affected by non-linearities. The main point of this paper is not to make precise forecasts for which certainly the lensing term should not be neglected, but to study the effect of non-linear corrections in the $C_\ell$'s coming from clustering and RSD.

In the next section we derive a `flat sky approximation' for density and RSD which is surprisingly accurate even at low $\ell$. In section~\ref{s:1loopPm} we describe and compare four different non-linear prescriptions for the power spectrum in redshift space, which can be found in the literature. This section is not new but we spell out these approximations for completeness. In Section~\ref{s:1loopCl} we compute the $C_\ell$'s from the different approximations and compare them with the linear and halofit results. We also compare our theoretical predictions to measurements made from a set of COLA $N$-body simulations. In Section~\ref{s:con} we discuss our findings and conclude.

\section{The flat sky approximation}
\begin{figure}[t]
    \centering
    \includegraphics[width=8cm,height=7cm]{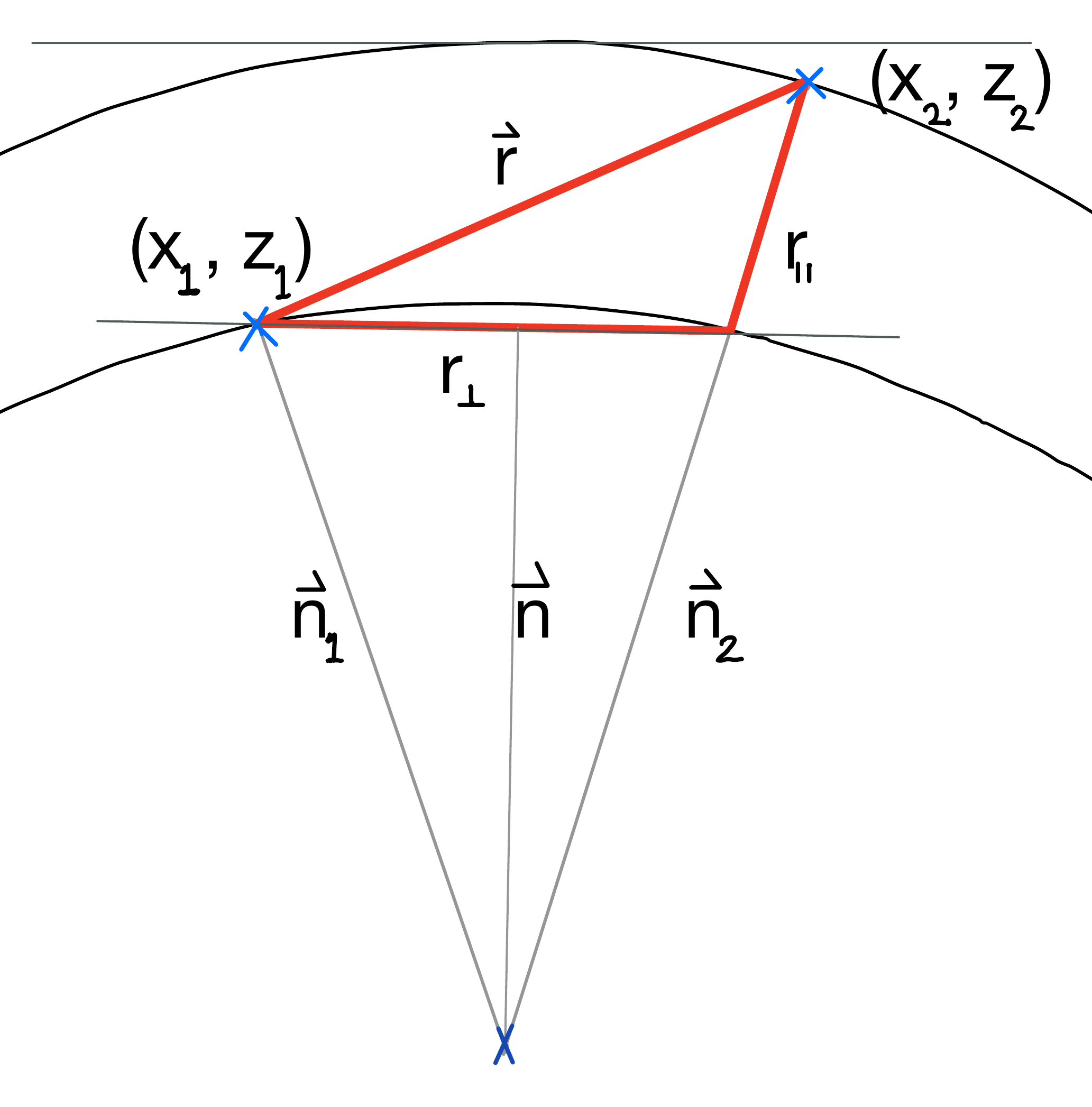}
    \caption{We show the positions $(\bx_1,z_1)$ and $(\bx_2,z_2)$ on the background lightcone of an observer situated at $X$ and their flat sky approximations.}
    \label{f:drawing}
\end{figure}

We want to compute the angular power spectrum $C_\ell(z_1, z_2)$ for galaxy number counts from the 3-dimensional power spectrum, where $z_1$ and $z_2$ are two (relatively close) redshifts. We start from the  correlation function in configuration space which in principle depends on two spatial positions and two redshifts, $\xi(\bx_1,z_1;\bx_2,z_2)$ where $(\bx_1,z_1)$ and $(\bx_2,z_2)$ are constrained to lie on our background lightcone. We assume that the redshifts are relatively close so that the time evolution between $z_1$ and $z_2$ can be neglected. Then the correlation function depends only on $\rvec=\bx_2-\bx_1$ and $\bar z=(z_1+z_2)/2$ (see Fig.~\ref{f:drawing}).  This correlation function in real space, $\xi(\rvec,\bar{z})$, is the Fourier transform of the power spectrum
\be \label{eq:xsi-Fourier}
\xi(\rvec,\bar{z}) = \frac{1}{(2\pi)^3} \int d^3 \kvec P(\kvec, \bar{z}) e^{-i \kvec \cdot \rvec  }.
\ee  
 Let us now consider the flat sky approximation, which amounts to assuming that the direction from the observer to the points $\bx_1$ and $\bx_2$ are nearly equal,  $\mathbf{n}_1\simeq {\mathbf{n}_2}= \mathbf{{n}}$, i.e. the survey covers a relatively small patch of the sky in a fixed direction $\bn$.
 This is the situation for which Eq.~(\ref{e:poldk}) can be used as an approximation for the power spectrum. In this case we 
can also decompose the separation vector $\rvec$  into components perpendicular and parallel to the line of sight direction $\mathbf{n}$, as shown in Fig.~\ref{f:drawing},
 so we have:
\bea \label{eq:r_decomposition}
&\rvec = \mathbf{r_{\perp}} + r_{\parallel} \textbf{n},\nonumber \\ 
& r_{\parallel} = r \nu \simeq \chi(z_2) - \chi(z_1) \simeq \frac{\Delta z}{H(\bar{z})} \, ,
\eea
where $\chi(z)$ is the comoving distance to redshift $z$, $\nu = \hat\rvec \cdot \hat{\mathbf{n}}$ as shown in Fig.~\ref{f:drawing}. Similarly in k-space we define
\begin{align} \label{eq:k_decomposition}
\kvec &= \mathbf{k_{\perp}} + k_{\parallel} \mathbf{n}, \nonumber \\
 k_{\parallel} & = k \mu = k \, \khat \cdot \mathbf{n} \, .
\end{align}
We now introduce the dimensionless two dimensional vector $\bell$ by $\mathbf{k_{\perp}} \equiv \bell/  \chi(\bar{z})$. Therefore by using Eqs.~(\ref{eq:r_decomposition}) and (\ref{eq:k_decomposition}) we can rewrite Eq.~(\ref{eq:xsi-Fourier}) as:
\be \label{eq:xsi-decomposed}
\xi(\rvec,\bar{z}) = \frac{1}{(2\pi)^3} \int \frac{d^2 \bell}{\chi^2(\bar z)} dk_{\parallel} P(\kvec, \bar{z}) e^{-i \left( \frac{\bell \cdot \mathbf{r_{\perp}} }{\chi(\bar{z})} +  k_{\parallel} \frac{(z_2-z_1)}{H(\bar{z})} \right) }.
\ee 
On the other hand, we know how to compute $\xi(\rvec,z)$ from the angular power spectrum. In the flat-sky approximation this yields (see e.g.~\cite{mybook})
\be \label{eq:flat-sky-usual}
\xi(\mathbf{r_{\perp}}, z_1, z_2 ) = \frac{1}{(2\pi)^2}\int d^2\bell\, C_\ell( z_1,z_2) e^{-i\bell\cdot \mathbf{r_{\perp}} /\chi(\bar{z}) }.
\ee 
By comparing Eqs.~(\ref{eq:xsi-decomposed}) and (\ref{eq:flat-sky-usual}), we find the relation between the angular power spectrum and the three dimensional power spectrum in Fourier space as
\be \label{eq:flat-sky-Ruth}
C_\ell( z_1,z_2) = \frac{1}{2 \pi \chi^2(\bar{z})}\int_{-\infty}^{+\infty} d k_{\parallel} 
P\left(k, 
\bar{z} \right) e^{-i k_{\parallel}(z_2-z_1)/H(\bar{z})}  \, , \ee
for $k=\sqrt{k_{\parallel}^2 + ({\ell}/{\chi})^2 }$.
Note that this approximation is not equivalent to the Limber approximation~\cite{Limber:1954zz} which is often used for weak lensing calculations where $k\simeq (\ell+1/2)/\chi(z)$ is used instead of an integration of the power spectrum times the Bessel function. In our flat sky approximation we identify the flat sky vectors 
\be
\mathbf{k_{\perp}} \equiv \bell/  \chi(\bar{z}),
\ee
and integrate over $k_\parallel$. More details about this approximation shall be discussed in a forthcoming paper~\cite{William}.
\begin{figure}[ht]
    \centering
    \includegraphics[scale=0.47]{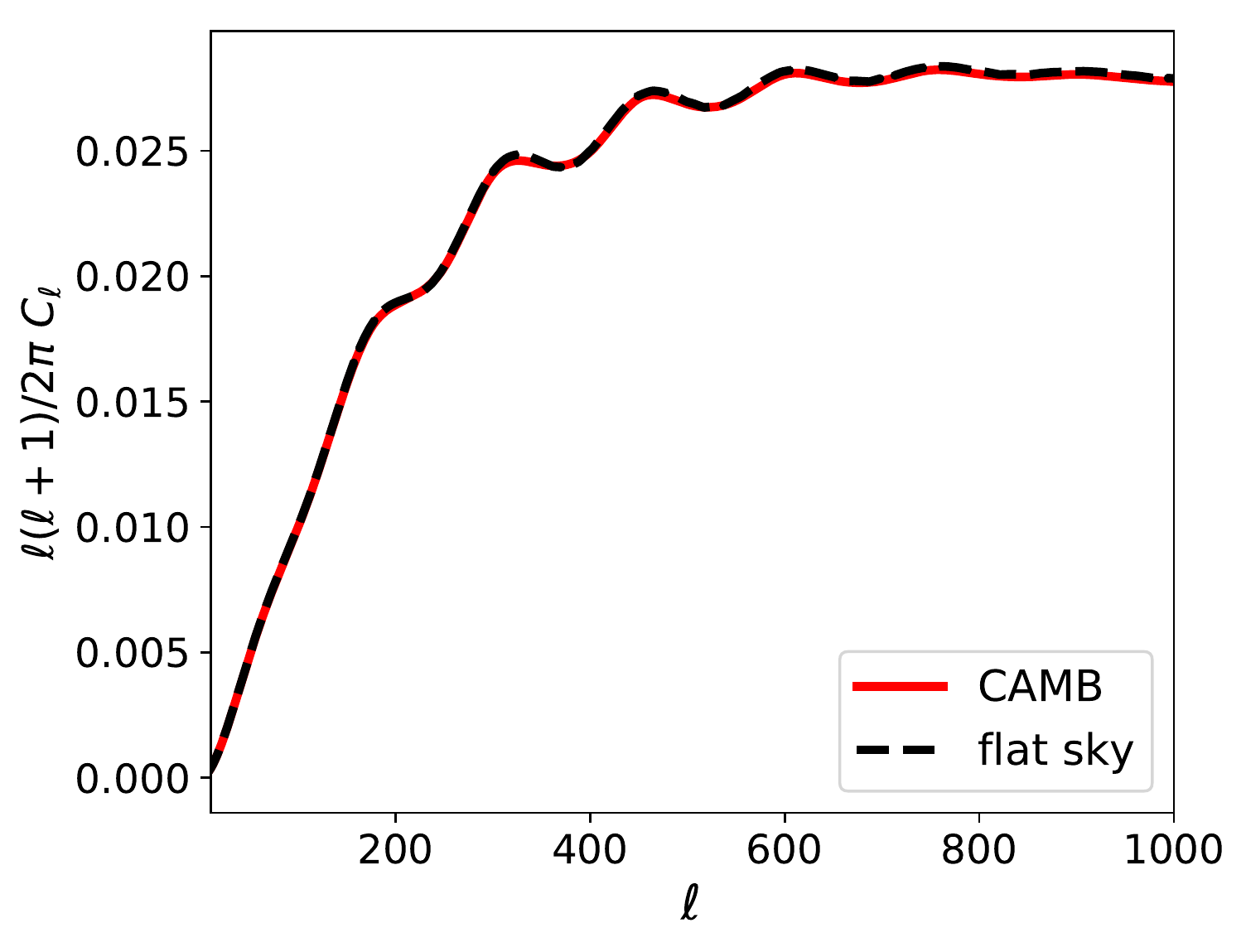}
    \includegraphics[scale=0.47]{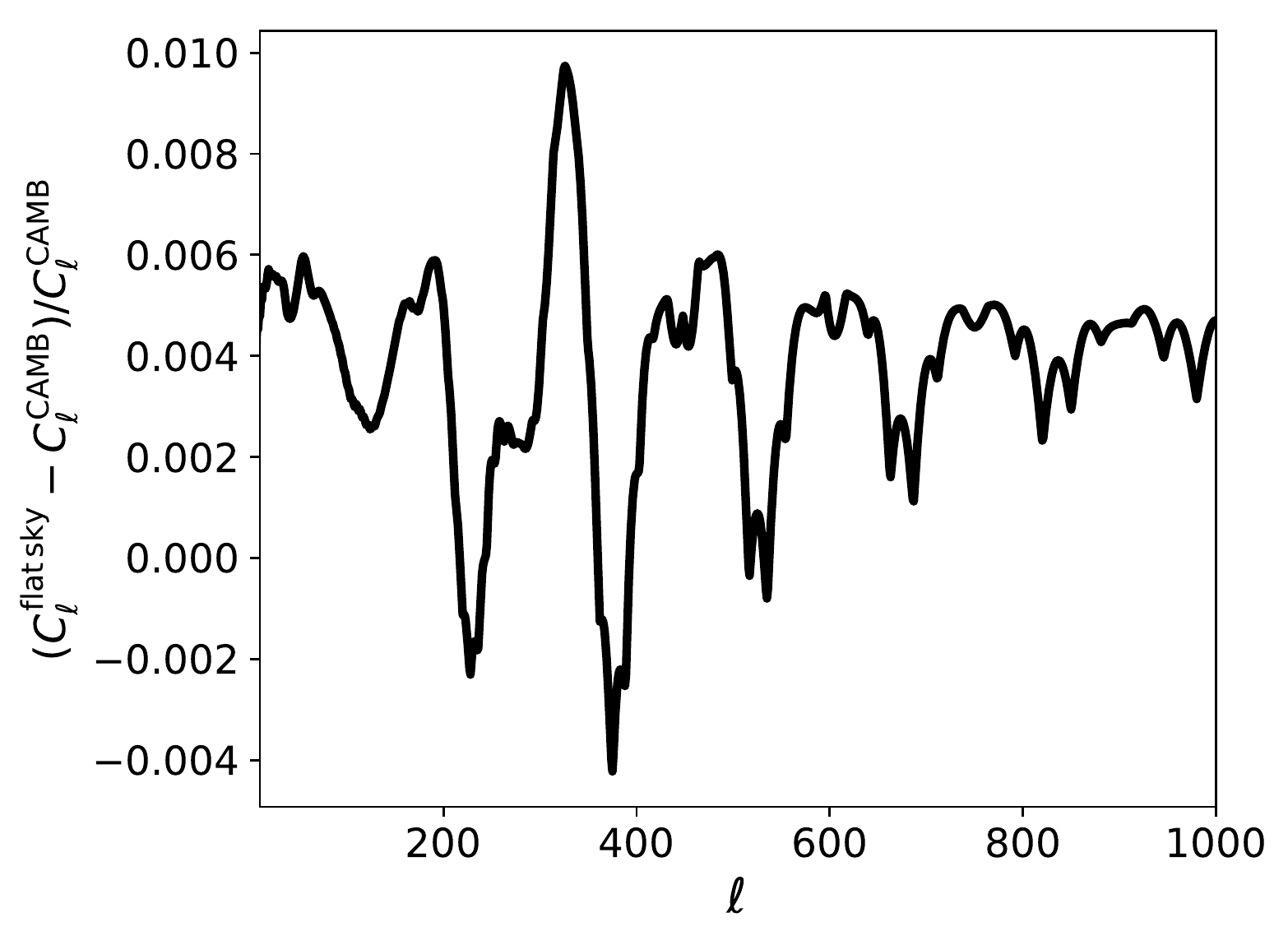}
    \caption{The left panel shows the comparison between the flat-sky approximation of Eq.~(\ref{eq:flat-sky-Ruth}) and the angular power spectrum computed by CAMB that uses Eq.~(\ref{eq:Pk_to_Cell_exact}) at $z=1$, using a top hat window function with $\Delta z = 0.1$. The right panel shows the relative difference between the two.}
    \label{fig:Cell_flat_sky_auto}
\end{figure}
Contrary to Limber's approximation, which is bad for the density and RSD contributions to number counts (see, e.g.~\cite{DiDio:2018unb}), this approximation turns out to be excellent for close redshifts $z_1 \simeq z_2$, when compared to the exact definition of $C_\ell( z_1,z_2)$ which, at low $\ell$, is given by (see Appendix B of \cite{Bonvin:2011bg}, where we have added the bias dependence)
\bea \nonumber
C_\ell( z_1,z_2) &=& \frac{2}{\pi} b(z_1)b(z_2)\int dk \, k^2 P_m(k,z_1,z_2) \left[ j_\ell(k\, \chi(z_1)) j_\ell(k\, \chi(z_2)) \right. \\
&&\left. - \frac{f(z_2)}{b(z_2)} j_\ell(k\, \chi(z_1)) j''_\ell(k\, \chi(z_2)) +  \frac{f(z_1)}{b(z_1)} \frac{f(z_2)}{b(z_2)} j''_\ell(k\, \chi(z_1)) j''_\ell(k\, \chi(z_2))\right]. \label{eq:Pk_to_Cell_exact}
\eea 
Here $P_m(k,z_1,z_2)$ is the matter power spectrum and $b(z_1)$, $b(z_2)$ are the linear tracer biases at $z_1$ and $z_2$.
For large $\Delta z$ the flat sky approximation gets worse. 
This is because this approximation corresponds to replacing the spherical Bessel function by their lowest frequency modes, assuming that $|\chi(z_1)-\chi(z_2)|\ll \chi(z_1),~\chi(z_2)$ which is no longer valid when the redshift difference becomes large (see~\cite{William} for a detailed derivation -- for large redshift differences we additionally need to model the decoherence in $P(k,z_1,z_2)$ correctly, e.g.\ with the fitting function of \cite{Chisari:2019tig}).
In Fig.~\ref{fig:Cell_flat_sky_auto} we compare the angular power spectrum for $z_1 = z_2 = 1$ computed in the flat sky approximation Eq.~\eqref{eq:flat-sky-Ruth} with the one computed with the exact formula of Eq.~(\ref{eq:Pk_to_Cell_exact}). The differences are at most $1 \%$.
\section{Non-linear corrections to the power spectrum in redshift space}\label{s:1loopPm}
In this section we give a summary of different non-linear corrections to the power spectrum that can be found in the literature. More precisely we consider four different approaches: 1-loop corrections from standard Newtonian (Eulerian) perturbation theory which we shall term SPT, 1-loop corrections from Lagrangian perturbation theory (LPT), corrections from effective field theory of large scale structure (EFT) and the Taruya-Nishimichi-Saito (TNS) model. Some important references for each of these approaches are \cite{Heavens:1998es,Matsubara:2007wj,delaBella:2018fdb,Taruya:2010mx} respectively. We perform all perturbative calculations at the one-loop level (see Appendix~\ref{Appendix:A} for details). %
We also make use of a set of measurements of the redshift power spectrum from COLA simulations. These represent our most accurate prediction, with which we can compare the perturbative approaches. These simulations are described briefly below. At the end of this section we  compare the different approximations to these simulations.
\subsection{COLA}
We have run a set of 10 Parallel COmoving Lagrangian Acceleration (PICOLA) simulations \citep{Howlett:2015hfa,Winther:2017jof} of box size $1024 \, \mbox{Mpc}/h$ with $1024^3$ dark matter particles and a starting redshift $z_{\rm ini}=49$. These are all run  under a similar $\Lambda$CDM cosmology with Planck parameters~\citep{Planck:2015xua}: $\Omega_m = 0.315$, $\Omega_b=0.0493$, $h=0.674$, $n_s=0.965$ and $\sigma_8(z=0) = 0.811$.
The simulation redshift space power spectrum multipoles are measured using the distant-observer (or flat sky) approximation\footnote{That is, we assume the observer is located at a distance much greater then the box size ($r\gg 1024 \, \mbox{Mpc}/h$), and so all lines of sight are treated as being parallel to the chosen Cartesian axes of the simulation box. Next, we disturb the position of the matter particles using their velocity components ($v_x, v_y$ or $v_z$).} and are then averaged over three line-of-sight directions. We further average over the 10 PICOLA simulations. We measure the first three multipoles, the monopole, quadrupole and hexadecapole. Using these we can then construct the full anisotropic power spectrum, $P(k,\mu)$\footnote{Note that the hexadecapole at the redshifts considered here is already very small in magnitude and so the exclusion of higher order multipoles will only negligibly affect the form of $P(k,\mu)$.}
\begin{eqnarray}
P^{\rm sCOLA}_{\rm tot}(\textbf{k},z) &=& 2P_0(k,z) + \frac{2}{5}\mathcal{L}_2(\mu)P_2(k,z)\nonumber \\ &+& \frac{2}{9}\mathcal{L}_4(\mu) P_4(k,z) + {\rm [higher \ order \ multipoles]},
\label{colapkmu}
\end{eqnarray}
where $\mathcal{L}_i$ is the Legendre polynomial of order `i' and $P_i$ is the $i^{\rm th}$ multipole which is an average over the measurements made from the COLA simulations. Finally, we note that the COLA method is an approximate method and has been shown to deviate from the full N-body approach at smaller scales \cite{Izard:2015dja,Bose:2019ywu,Blot:2018oxk}. This issue can be ignored as we simply use these simulations as a benchmark in accuracy with which to compare the less accurate perturbative predictions outlined next. For example, the redshift space monopole for lowly biased halos was shown to be accurate to full N-body to within a few percent at $z\leq1$ up to $k=0.7h/{\rm Mpc}$ in \cite{Izard:2015dja}. On the other hand, the quadrupole deviates by up to $10\%$ at $z=1$ at $k=0.7h/{\rm Mpc}$ in the same paper. Regarding this issue, we expect the dark matter monopole and quadrupole to perform better than the halo multipoles, and for their accuracy to improve at higher redshifts. Furthermore, we expect the theoretical models discussed in this section to perform significantly  worse at these scales.

\subsection{SPT}
One-loop contributions to the power spectrum in redshift space (denoted by superscript $s$) in the context of SPT  are already well-established and have previously been calculated in the literature (for a review see e.g. \cite{Bernardeau:2001qr}). Here as a reference we point to Eq.~(15) of \cite{Heavens:1998es}: 
\begin{align}\label{eq:ptot_SPT}
    P^{\rm sSPT}_{\rm tot}(\textbf{k},z)&\equiv P^s_{\rm lin}+ P^{\rm sSPT}_{\rm 1-loop}= P^s_{\rm lin}+P^s_{22}+P^s_{13}\\ \nonumber
    &= (1+\beta\mu^2)^2b^2P_{\rm lin}(k,z)+2\int\frac{d^3\textbf{q}}{(2\pi)^3}P_{\rm lin}(q , z)P_{\rm lin}(|\textbf{k-q}|,z )[F_2^S(\textbf{q},\textbf{k-q})]^2\\ \nonumber
    &\quad +6(1+\beta\mu^2)b P_{\rm lin}(k,z)\int \frac{d^3\textbf{q}}{(2\pi)^3}P_{\rm lin}(q,z) F_3^S(\textbf{q},\textbf{-q},\textbf{k}) \, ,
\end{align}
where $P_{\rm lin}(k,z)$ is the linear power spectrum in real space, $\beta\equiv f/b$, $f$ being the linear growth rate and $b$ being the linear bias, $\mu = \hat{\kvec} \cdot \mathbf{n}$, and  $F_2^S(\textbf{q},\textbf{k-q})$ and $F_3^S(\textbf{q,-q,k})$ are the kernels of higher order perturbations. Their expressions are computed from Eq. (13) of Ref.~\cite{Heavens:1998es} by neglecting higher order biases. The details of the integrations that appear in Eq.~(\ref{eq:ptot_SPT}) are given in Appendix~\ref{Appendix:A} for completeness. Further, since we only consider dark matter we set $b=1$.

\subsection{LPT}
The power spectrum using Lagrangian perturbation theory (LPT) is given in Eq.~(63) of Ref.~\cite{Matsubara:2007wj},
\begin{align}\label{eq:pLPT}
&  P^{\rm sLPT}_{\rm tot}(\mathbf{k},z) =
  \exp\left\{ - k^2[1 + f(f+2)\mu^2] A \right\}
\nonumber\\
& \quad \times
  \left\{ P^{\rm sSPT}_{\rm tot}(\textbf{k},z )
      +\, (1 + f\mu^2)^2[1 + f(f+2)\mu^2] k^2 P_{\rm lin}(k,z) A
  \right\},
\end{align}
where 
\begin{equation}
  A = \frac{1}{6\pi^2} \int dq P_{\rm lin}(q,z). 
\end{equation}
The pre-factor encodes a damping on small scales from velocity dispersion.
\subsection{EFT}
We also consider effective field theory of large scale structure \cite{Baumann:2010tm,Carrasco:2012cv,Senatore:2014vja,Lewandowski:2015ziq,Perko:2016puo,Foreman:2015lca} where counter terms are added to the SPT power spectrum, for which we  refer to Eq. (3.8) of Ref.~\cite{delaBella:2018fdb}
\begin{equation}\label{eq:pEFT}
     P^{\rm sEFT}_{\rm tot}(\mathbf{k},z)
    =  P^{\rm sSPT}_{\rm tot}(\mathbf{k},z)
    -
    2 \sum_{n=0}^3
    c_{2|\delta_s,2n} \mu^{2n} \frac{k^2}{k_{\rm nl}^2} P_{\rm lin}(k,z),
\end{equation}
where
\begin{subequations}
    \begin{align}
        \label{eq:ctrterm-condition-a}
        c_{2|\delta_s,6} & = f^3 c_{2|\delta_s,0} - f^2 c_{2|\delta_s,2} + f c_{2|\delta_s,4} , \\
        \label{eq:ctrterm-condition-b}
       c_{2|\delta_s,8} & = 0 .
    \end{align}
\end{subequations}
We do not apply a resummation scheme as is commonly done in the literature.  The effect of resummation has been shown to have a low impact on the fitting to COLA data conducted in \cite{delaBella:2018fdb}. The values of the counter term coefficients have been determined by fitting to the COLA simulations. This follows a similar procedure to \cite{Bose:2019psj}. We refer the reader to this work for justifications and details of this procedure. This is briefly described in Appendix~\ref{Appendix:fit} where also the numerical values of the fitting parameters are given.

\subsection{TNS}
The last model we consider is the TNS model. This model was introduced in \cite{Taruya:2010mx} and is one of the best approaches to perturbation theory known at present, having been applied in the recent BOSS galaxy clustering analysis \citep{Beutler:2013yhm,Beutler:2016arn}. It has also been thoroughly validated against simulations and has stood up to other perturbative models \cite{Nishimichi:2011jm,Taruya:2013my,Ishikawa:2013aea,Zheng:2016zxc,Gil-Marin:2015sqa,Gil-Marin:2015nqa,Bose:2017myh,Bose:2016qun,Markovic:2019sva,Bose:2019ywu,Bose:2019psj}. The model is given by \cite{Taruya:2010mx}
 \begin{align}
 P^{sTNS}_{\rm tot}(k,z) = \frac{1}{1 + (k^2\mu^2 \sigma_v^2)/2}  \Big[& P^{\delta \delta}_{\rm 1-loop} (k,z)  + 2 \mu^2 P^{\delta \theta}_{\rm 1-loop}(k,z) +  \mu^4 P^{\theta \theta}_{\rm 1-loop} (k,z) \nonumber \\ &+ A(k,\mu,z) +  B(k,\mu,z) + C(k,\mu,z)  \Big].
 \label{eq:pTNS}
 \end{align} 
The terms in brackets are all constructed within SPT, with $\delta \delta$, $\delta \theta$ and $\theta \theta$ denoting density-density, density-velocity and velocity-velocity 1-loop power spectra. The perturbative correction terms $A,B$ and $C$ are non-linear corrections coming from the RSD modelling while the prefactor is added for phenomenological modeling of the Fingers of God effect. Within this prefactor, $\sigma_v$, is a free parameter that is fit to the COLA simulations (see Appendix~\ref{Appendix:fit}). We refer the reader to \cite{Taruya:2010mx,Bose:2019psj} for a detailed description of the  components $A$, $B$ and $C$ of the model but we give some basic expressions in Appendix~\ref{Appendix:ABC}.  

\subsection{Comparisons}
\begin{figure}[H]
    \begin{center}
    \includegraphics[scale=0.43]{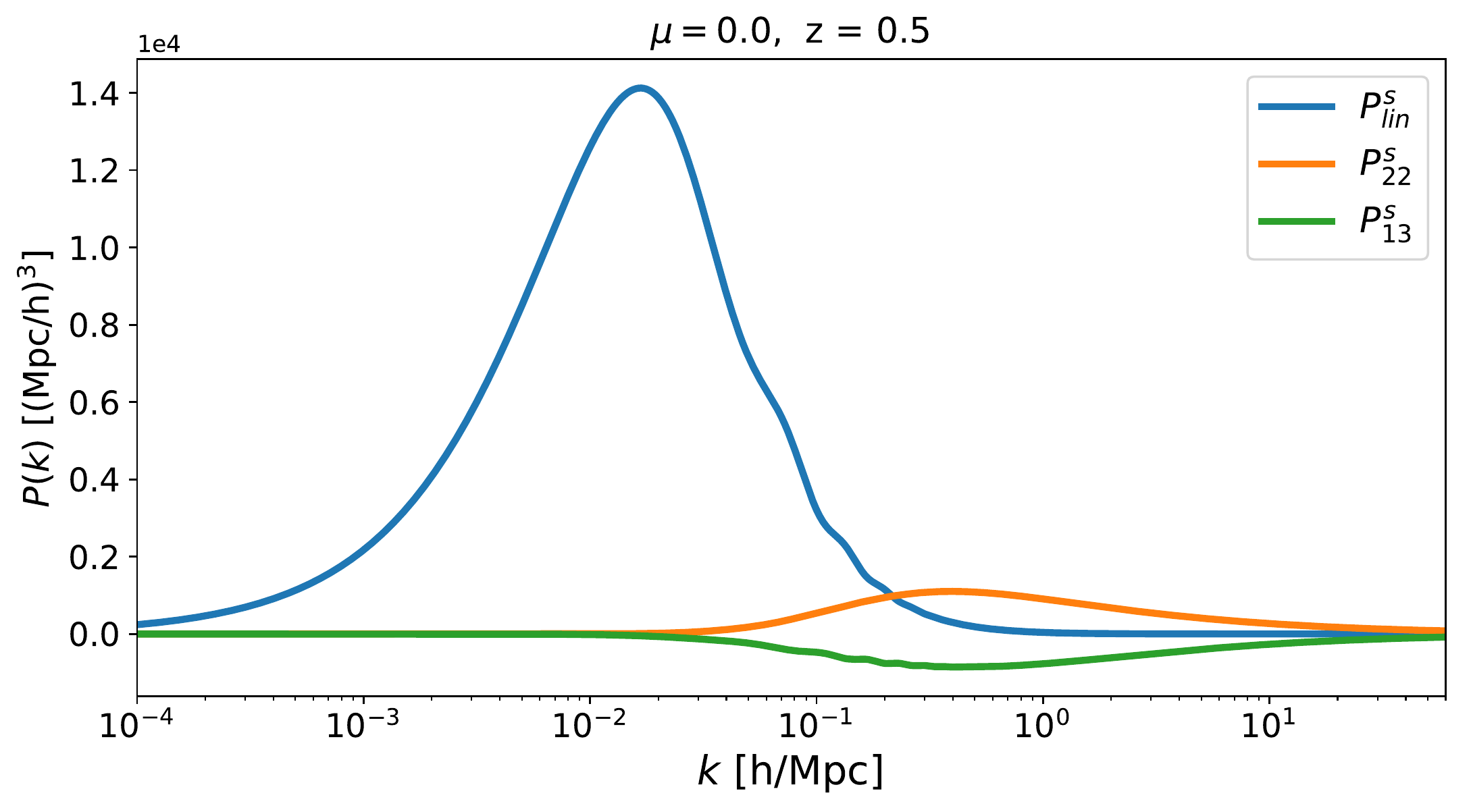}\hspace*{3.5cm}~~  \\
    \includegraphics[scale=0.43]{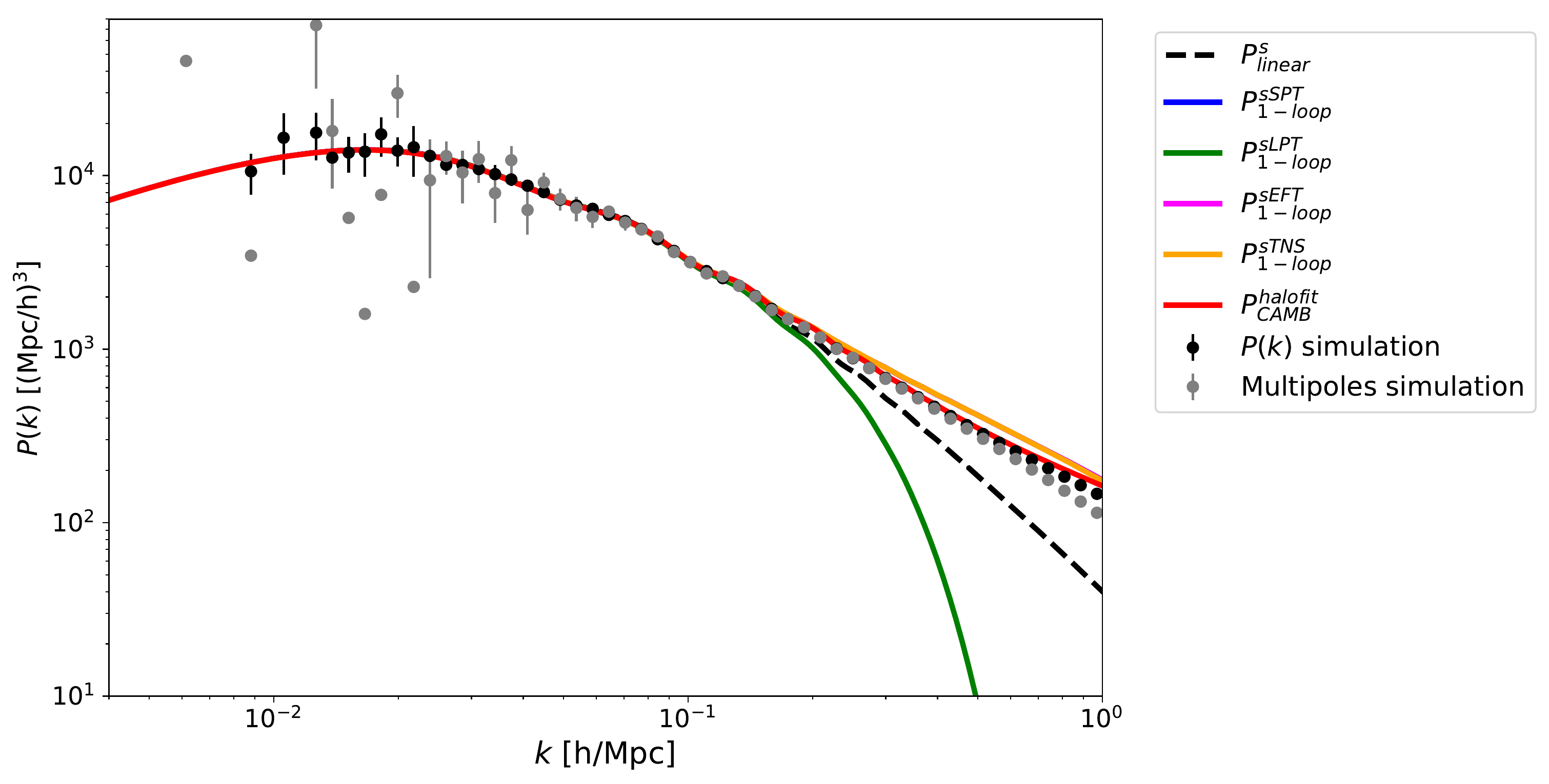}\\
    \includegraphics[scale=0.43]{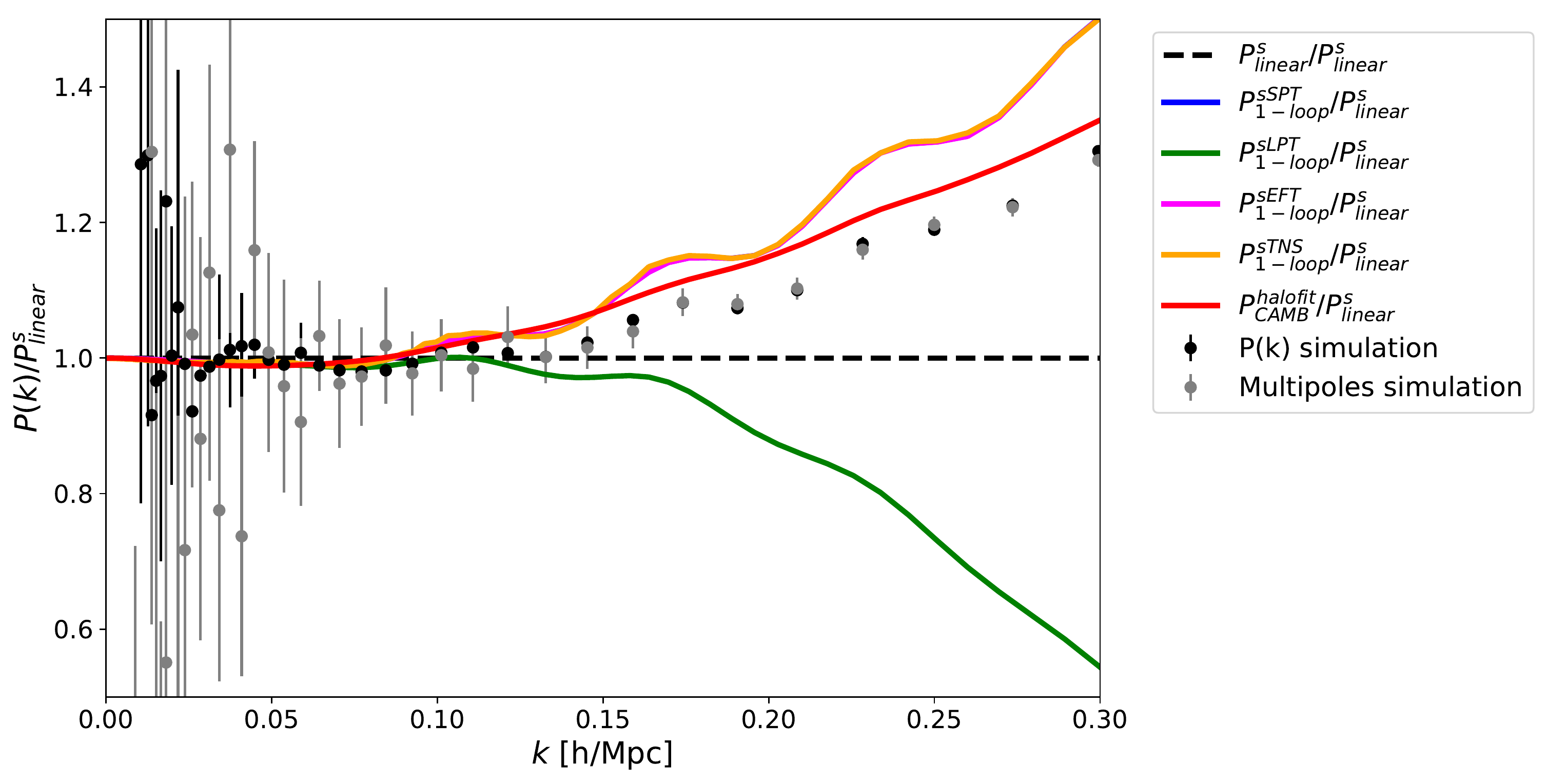}
    \end{center}
    \caption{The power spectrum $P(k)$ in redshift space in the transversal direction, $\mu=0$ (effectively the density power spectrum), with $b=1$ and $z = 0.5$. The upper panel shows the linear spectrum $P^s_{11}$ (blue) along with the one-loop contributions $P^s_{22}$ (orange) and $P^s_{13}$ (green). The middle panel shows the comparison between $P^{\rm sSPT}_{\rm 1-loop}$, $P^{\rm sLPT}_{\rm 1-loop}$, and $P^{\rm sEFT}_{\rm 1-loop}$ and $P^{\rm sTNS}_{\rm 1-loop}$ defined respectively in Eq.~(\ref{eq:ptot_SPT}), Eq.~(\ref{eq:pLPT}), Eq.~(\ref{eq:pEFT}) and Eq.~(\ref{eq:pTNS}). In the lower panel the ratios of the corresponding non-linear spectra and the linear one are shown. The black dots show the monopole of the $N$-body simulations while the grey dots also include the quadrupole and the hexadecapole available from COLA.
    \label{fig:Pk_comparison0}}
\end{figure}

\begin{figure}[H]
    \begin{center}
    \includegraphics[scale=0.43]{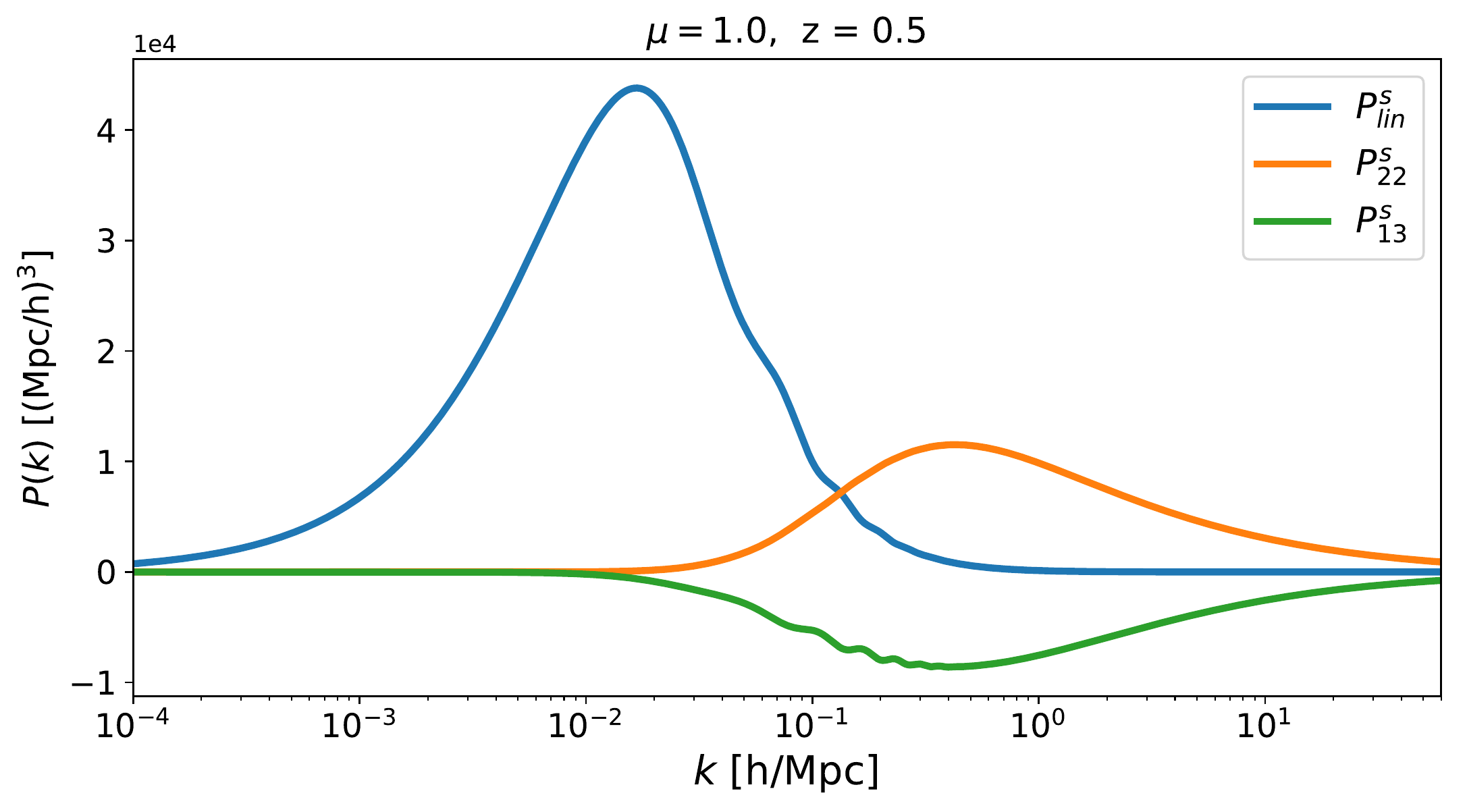}\hspace*{3.5cm}~~  \\
    \includegraphics[scale=0.43]{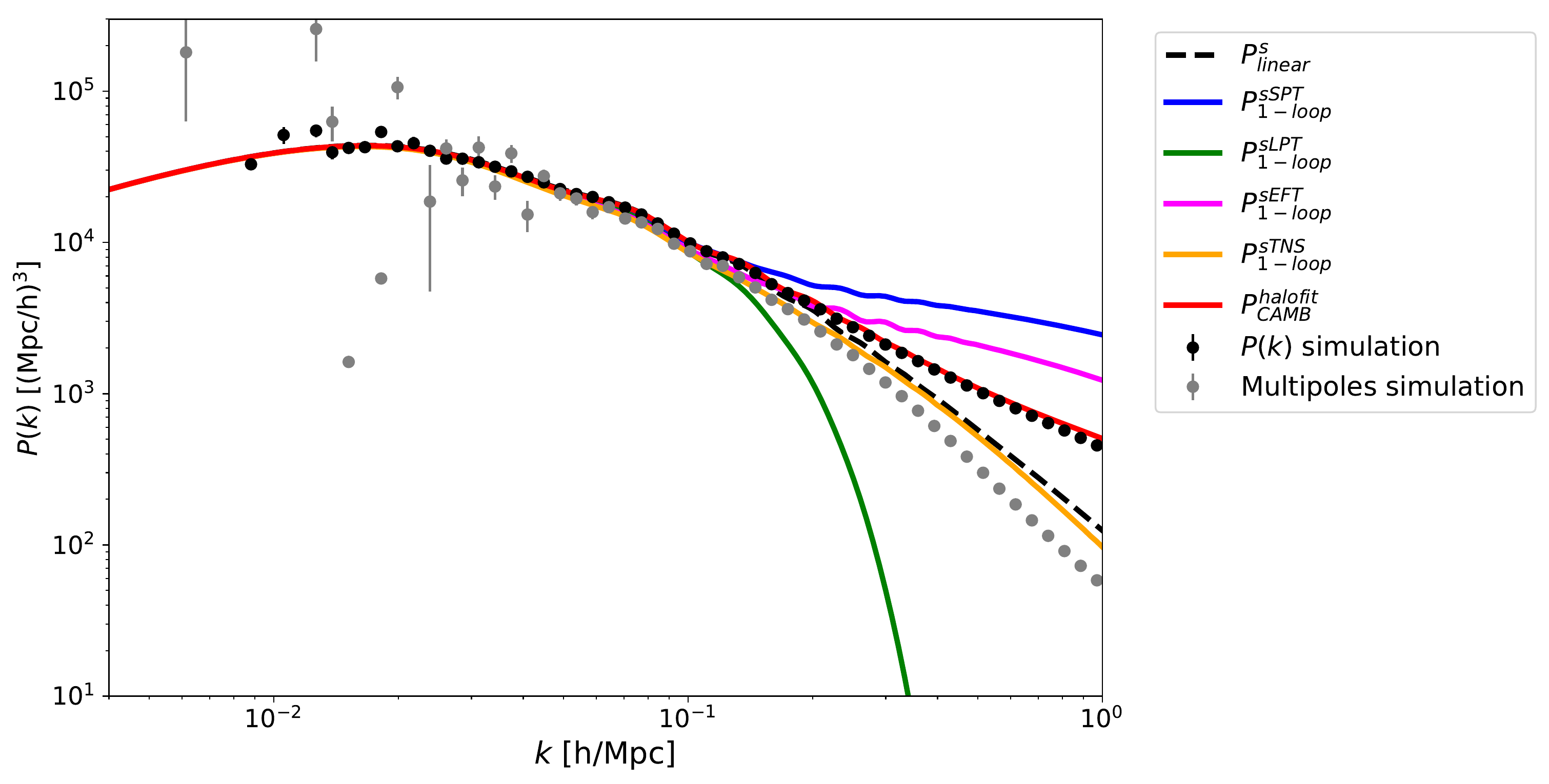}\\
    \includegraphics[scale=0.43]{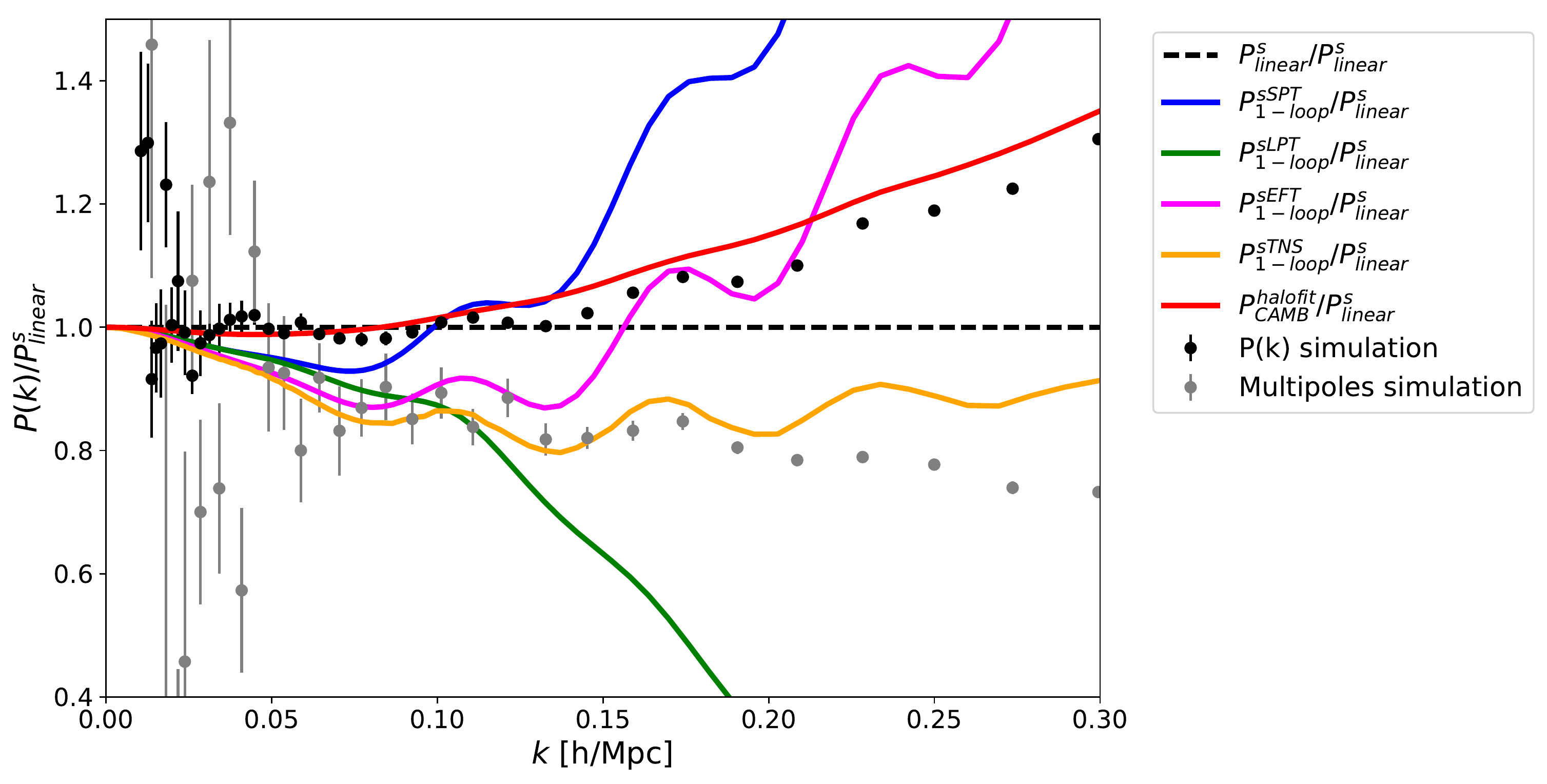}
    \end{center}
    \caption{The same as Fig.~\ref{fig:Pk_comparison0} but in the radial direction, $\mu=1$, where redshift space distortions are important. We see that the Kaiser formula used for halofit and for the black COLA points does not provide a good fit to the RSD even at mildly non-linear scales.
    \label{fig:Pk_comparison}}
\end{figure}

In this section we compare Eq.~(\ref{eq:ptot_SPT}) (SPT - blue), Eq.~(\ref{eq:pLPT}) (LPT - green), Eq.~(\ref{eq:pEFT}) (EFT - magenta) and Eq.~(\ref{eq:pTNS}) (TNS - orange) with Eq.~(\ref{colapkmu}) (COLA reconstructed 2d spectrum - grey dots). We also compare Eq.~(\ref{e:poldk}) with $P_m(k)$ given by linear theory (linear Kaiser - dashed black), non-linear halofit power spectrum \cite{Takahashi:2012em} (halofit - red) and the matter power spectrum as measured from the simulations (black dots). These comparisons are done at $z=0.5$ and are shown in Fig.~\ref{fig:Pk_comparison0} and Fig.~\ref{fig:Pk_comparison} for $\mu=0$ (transversal direction) and $\mu=1$ (radial direction) respectively. We expect that the grey dots marking the reconstructed COLA 2D spectrum of Eq.~(\ref{colapkmu}) provide the most accurate modeling for the full spectrum.
This will be our benchmark for accuracy. 

We also show the one-loop contributions to $P_{\rm lin}$, $P_{13}$ and $P_{22}$, of Eq.~(\ref{eq:ptot_SPT}) in the upper panels of Fig.~\ref{fig:Pk_comparison0} and Fig.~\ref{fig:Pk_comparison}. They start to become important at $k \sim 0.1h$/Mpc for $\mu=0$, which is well known from the literature, and on smaller scales for $\mu=1$ as we can see in Fig.\ \ref{fig:Pk_comparison} and as we will also discuss later. Furthermore, $P_{13}$ and $P_{22}$ have opposite signs and their amplitudes are individually much larger than their sum, which is an indication for the well known bad convergence properties of SPT \cite{Carlson:2009it}. 

The 1-loop SPT power spectrum (blue line) at $z=0.5$ is shown in the middle panel of Fig.~\ref{fig:Pk_comparison} for $\mu=1$ and in  Fig.~\ref{fig:Pk_comparison0} for $\mu=0$ (in this plot the blue line is covered by the orange line). One sees clearly that SPT has too much power at small scales and fits the COLA simulations (grey dots) in a satisfactory way  only for $k\lesssim 0.1h/$Mpc.

The black dots are the COLA matter power spectrum multiplied by the `Kaiser factor' $(1+\beta\mu^2)^2$. They are accurate until about $k=0.13h/$Mpc. The keen reader may ask why the grey dots and black dots do not overlap in Fig.~\ref{fig:Pk_comparison0} at small scales. This could be due to inaccuracies in the COLA velocities used in computing the multipoles as well as the exclusion of higher order multipoles in Eq.~(\ref{colapkmu}).

Next we consider LPT (green curves). It is clear from Fig.~\ref{fig:Pk_comparison0} and Fig.~\ref{fig:Pk_comparison}, that the damping introduced in the LPT model is much too strong. Nevertheless, this correction can fit the power spectrum roughly until $k<0.13h/$Mpc which is already better than the SPT fit.

The EFT power spectrum is plotted as the magenta line (in Fig.~\ref{fig:Pk_comparison0} this line is covered by the orange line). Somewhat surprisingly, this fit is only a little but not significantly better than LPT for the angular scales considered. It represents a reasonable approximation until $k\simeq 0.15h/$Mpc. One key reason for the poor fit at $\mu=1$ is the lack of damping within the SPT spectrum which the EFT counter terms cannot suppress efficiently. The inclusion of resummation is expected to improve the fit (see for example \cite{Bose:2019psj}) but we leave this to future work.

Lastly, the TNS model is shown in orange. Clearly, this model represents the best fit to the full reconstructed simulated power spectrum for $\mu=1$ (compare the orange line and the grey dots in Fig.~\ref{fig:Pk_comparison}, lower panel). It can be used roughly until $k\simeq 0.2h/$Mpc.  
This is somewhat disappointing, as we aspire to achieve a good fit until $k\simeq 1h/$Mpc -- to reach convergence in the $C_\ell$ integral for narrow redshift bins we find that we need to go even to $2 h/$Mpc. On even smaller scales, corrections from  baryonic physics, that are not present in the simulations used here, can at any rate no longer be ignored.

We also note that for $\mu=0$ the SPT, EFT and TNS power spectra are identical, i.e. in Fig.~\ref{fig:Pk_comparison0} the blue, magenta and orange lines overlay. These spectra only differ in their treatment of redshift space distortions which are absent in the transversal direction, $\mu=0$. 

In Fig.~\ref{fig:Pk_comparison0} and~\ref{fig:Pk_comparison} we also show the comparison of the COLA measurements with the halofit model multiplied by the Kaiser factor given in Eq.~(\ref{e:poldk}) (red curve). While this approximation is excellent when fitted to the COLA matter power spectrum, see Fig.~\ref{fig:Pk_comparison0}, it does not correctly model the redshift space distortions. Hence, the higher order RSD and the non-linearity in the continuity equation which is not taken into account in this formula is very relevant. This is also clear from comparing the black dots, obtained from the matter power spectrum of the COLA simulations by multiplication with the Kaiser term, and the grey dots which represent the full sum of the simulated multipoles. It is also interesting to note that while the matter power spectrum of the simulations on small scales is larger than the linear power spectrum, adding all the multipoles actually reduces the power spectrum in redshift space on small scales when compared to the linear power spectrum. While the LPT approximation exaggerates this reduction of power, all other approximations either cannot model it at all or (in the case of TNS) underestimate this effect.  This is most visible in radial direction, $\mu=1$. In the transversal direction, $\mu=0$, the non-linear corrections from SPT, EFT and TNS all overshoot significantly while LPT is still too small. Here halofit provides the best approximation, see Fig.~\ref{fig:Pk_comparison0}. In the radial direction, $\mu=1$, only TNS manages to provide a reasonable fit for $k\gtrsim 0.1h$/Mpc, but for $k\gtrsim 0.2h$/Mpc it also starts to over-estimate the power significantly so that there is effectively no good analytical prescription available to model the redshift space power spectrum into the non-linear regime.

\FloatBarrier
\section{Non-linear correction to the angular power spectrum}\label{s:1loopCl}
To profit optimally from future galaxy redshift surveys (Euclid, DESI, 4MOST, SKA, ...) \cite{Amendola:2016saw,4MOST:2019,Aghamousa:2016zmz,Santos:2015bsa} we must also be able to model scales where non-linearities become relevant. Since the angular power spectrum is the true observable quantity, it is important to study the effects of non-linearities directly on this quantity.  In this section, we discuss the effect of non-linearities on the angular power spectrum using the different approaches discussed in the previous section to model them, and we study their effects at different redshifts and for different widths of the redshift bins considered.

\begin{figure}[h!]
    \centering
    \includegraphics[scale=0.5]{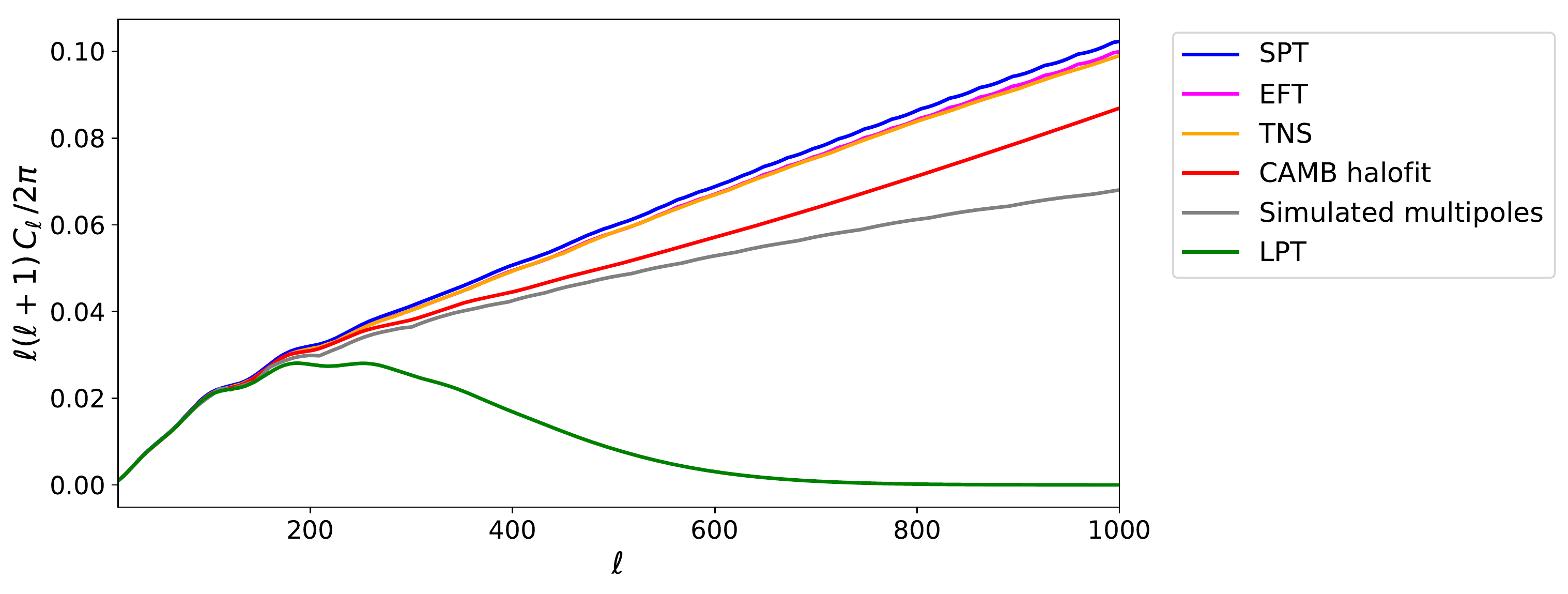}\\
    \includegraphics[scale=0.5]{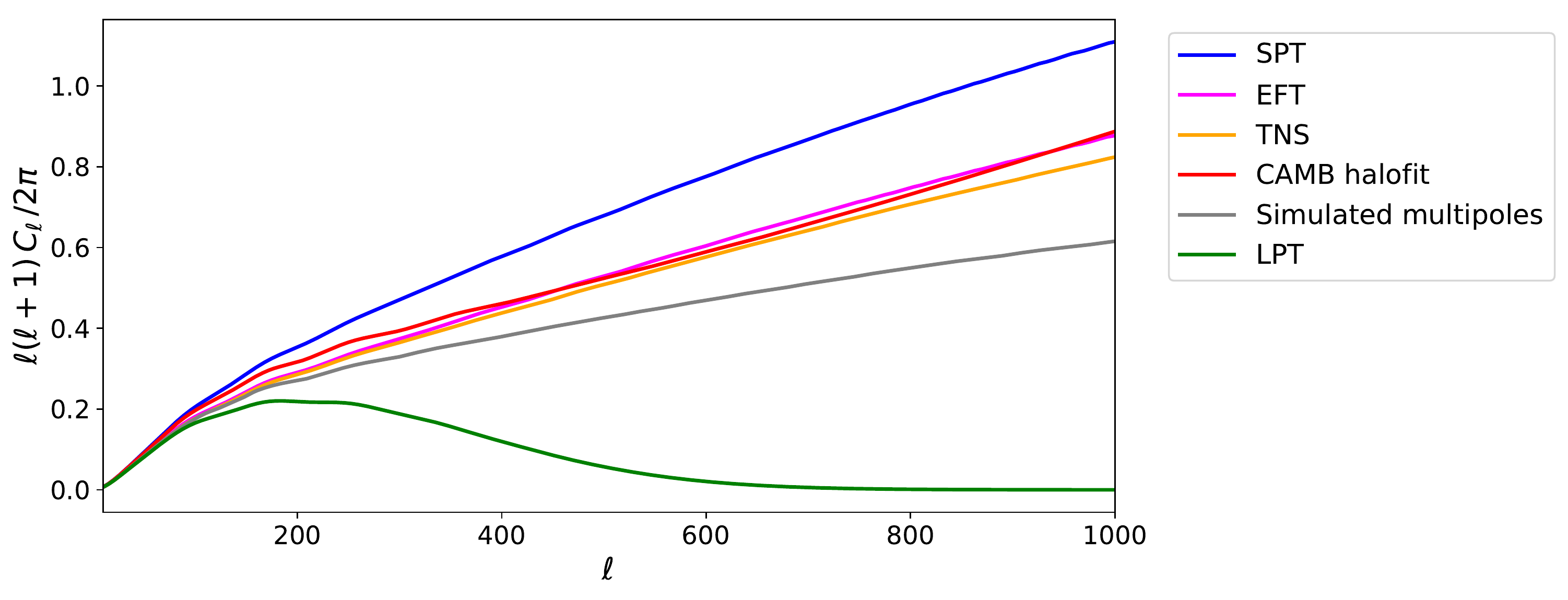}
    \includegraphics[scale=0.5]{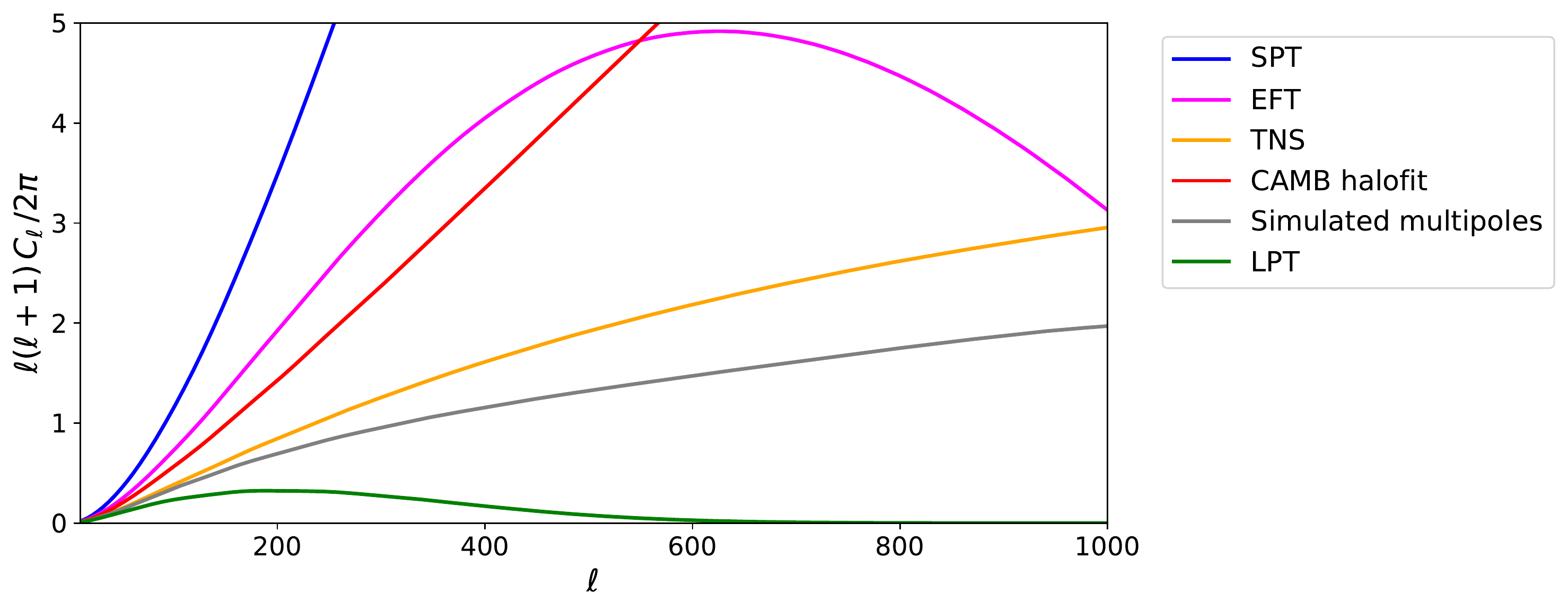}
    \caption{The plot shows the $C_{\ell}$'s from density and RSD at $z=0.5$ computed using the different approaches discussed in the text: $P^{\rm sSPT}_{\rm 1-loop}$, Eq.~(\ref{eq:ptot_SPT}), $P^{\rm sLPT}_{\rm 1-loop}$, Eq.~(\ref{eq:pLPT}), $P^{\rm sEFT}_{\rm 1-loop}$, Eq.~(\ref{eq:pEFT}), and $P^{\rm sTNS}_{\rm 1-loop}$,  Eq.~(\ref{eq:pTNS}), as well as the non-linear $C_{\ell}$ computed by {\sc CAMB} using the halofit model, and the simulated multipoles from COLA. The redshift bin width is $\De z=0.1$ for the top panel, $\De z = 0.01$ for the middle panel and $\De z = 0.001$ for the bottom panel. None of the models shown here manages to agree with the numerical simulations except on the largest scales and for wide redshift bins. }
    \label{fig:Cell_compare}
\end{figure}

Although the power spectrum $P(k,z)$ in  Fourier space, and its counterpart, the correlation function $\xi(r,\mu,z)$, provide some insight into galaxy observations on small scales, here we want to investigate how these non-linearities project onto the sky, i.e. onto the directly observable angular power spectrum. 

In the top panel of Fig.~\ref{fig:Cell_compare} we compare the $C_\ell$'s from the different non-linear approximations discussed in the previous section at redshift $z=0.5$ and using bin width $\De z=0.1$.
 For $\ell\lesssim 150$, which corresponds roughly to the non-linearity scale at $z=0.5$, the spectra agree relatively well. Beyond that scale they become very different, and even though in $k$-space TNS is a better approximation to the numerical results this is no longer true in $\ell$ space where the CAMB halofit (red line) seems to best mimic the COLA result (grey line), but also this result is more than 20\% off at $\ell = 1000$ from the COLA simulation and a better approximation is certainly needed. 

When smaller bin widths are chosen, $\De z=0.01$ for the middle panel and $\De z=0.001$ for the lower panel, the difference between the approximations and the COLA simulations becomes even worse. For these bin widths more small scale power enters the $C_\ell$'s which not only increases their amplitude but also makes them more sensitive to the treatment of non-linearities. 

\begin{figure}[t]
    \centering
    \includegraphics[scale=0.47]{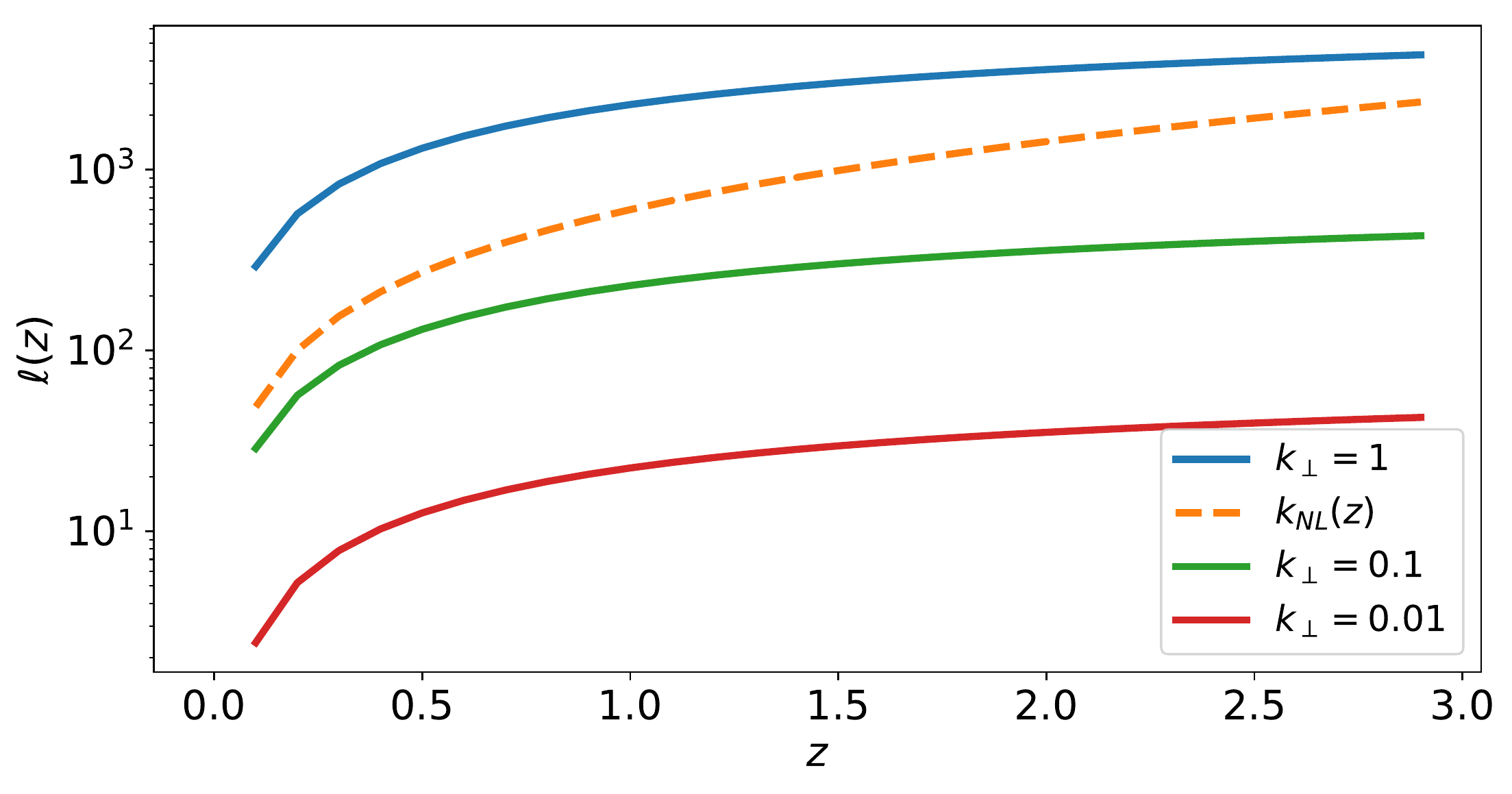}
    \caption{We show $\ell(k_\perp,z)$ for $k_\perp=(0.01\,,~0.1\,,~1) h/$Mpc as well as $\ell_{\rm NL}(z)=\ell(k_{\rm NL}(z),z)$ as a function of redshift $z$.}
    \label{fig:ellvsz2}
\end{figure}

We define the non-linearity scale through the condition
\be
\sigma(R_{\rm NL}) = 0.2
\ee
that was also used by Euclid \cite{Laureijs:2011gra,Rassat:2008ja}. Here $\sigma^2(R)$ is the usual variance of the mass fluctuation in a sphere of radius $R$,
\be
\sigma^2(R,z) \equiv \frac{1}{2\pi^2} \int_0^\infty \frac{dk}{k} \left( \frac{3 j_1(k R)}{k R} \right)^2 k^3 \delta^2(k,z) \, ,
\ee
so that $\sigma(R=8h/\mathrm{Mpc})=\sigma_8$. We then associate a non-linearity scale in Fourier space through
\be
k_{\rm NL}(z) = \frac{2\pi}{R_{\rm NL}(z)} \, .
\ee
A given transversal wave number $k_\perp$ at redshift $z$ roughly corresponds to a multipole
\be
\ell(k,z) \simeq k_\perp\chi(z) \,.
\ee
In Fig.~\ref{fig:ellvsz2} we show  $\ell(k_\perp,z)$ for three different values of $k_\perp$ as well as 
$\ell_{\rm NL}(z)=\ell(k_{\rm NL}(z),z)$.

\begin{figure}
    \centering
    \includegraphics[scale=0.45]{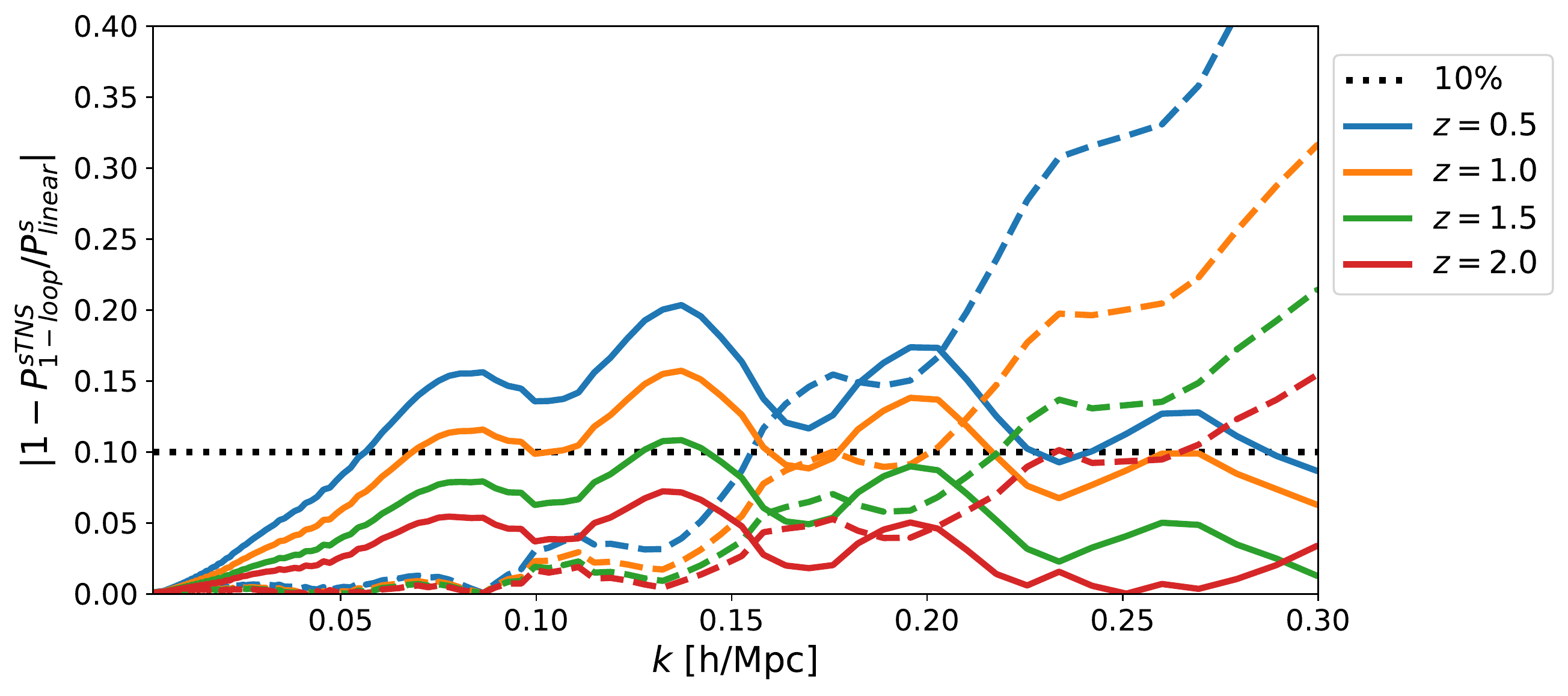}
    \caption{Relative differences between $P(k,\mu)^{\rm NL}$ and $P(k,\mu)^{\rm lin}$ for the TNS model for different redshifts. The dashed lines represent the density only ($\mu=0$), and the solid ones represent density plus RSD ($\mu =1$). The horizontal dotted black line is the $10 \%$ line. For the density-only spectrum non-linearities become important around $k=0.1 h$/Mpc, while in the $\mu=1$ spectrum with RSD the non-linearities appear on much larger scales.}
    \label{fig:reldiff_TNSk}
\end{figure}
In Fig.~\ref{fig:reldiff_TNSk} we compare linear and non-linear spectra for different redshifts for the TNS model in Fourier space. For $k<0.15h/$Mpc the density only spectra (dashed lines) are closer to the linear result than the spectra including RSD with $\mu=1$ (solid lines). This indicates that velocities exhibit non-linearities already on larger scales than the density. Roughly at $k=0.15h/$Mpc this trend is reversed. When we enter a more non-linear regime (after shell crossing), the velocities tend to damp the power in redshift space, so the density + RSD spectra are less non-linear than the density only spectra on these scales. Interestingly, the `cross-over' scale of $k=0.15h/$Mpc seems to be nearly redshift independent.
\begin{figure}[ht!]
    \centering
    \includegraphics[scale=0.5]{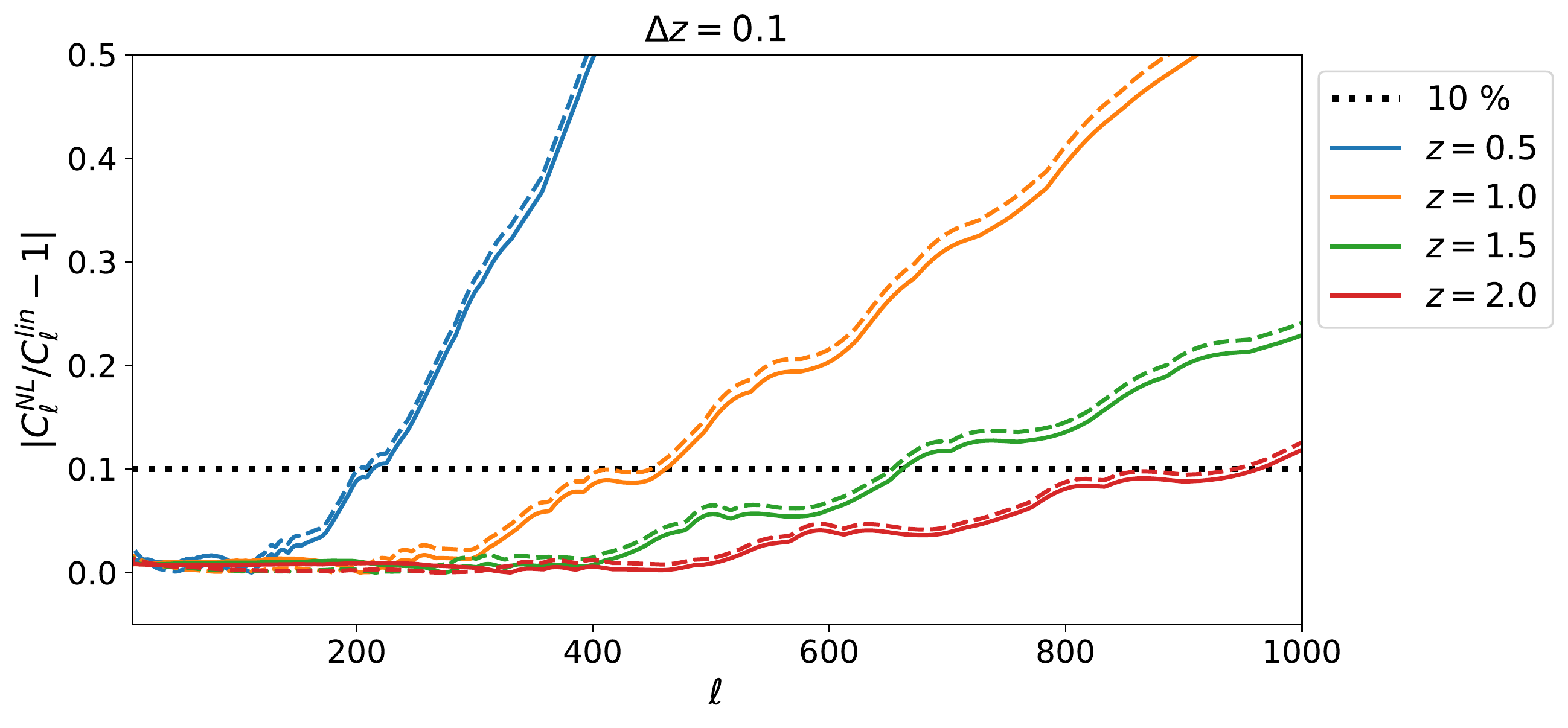}
    \includegraphics[scale=0.5]{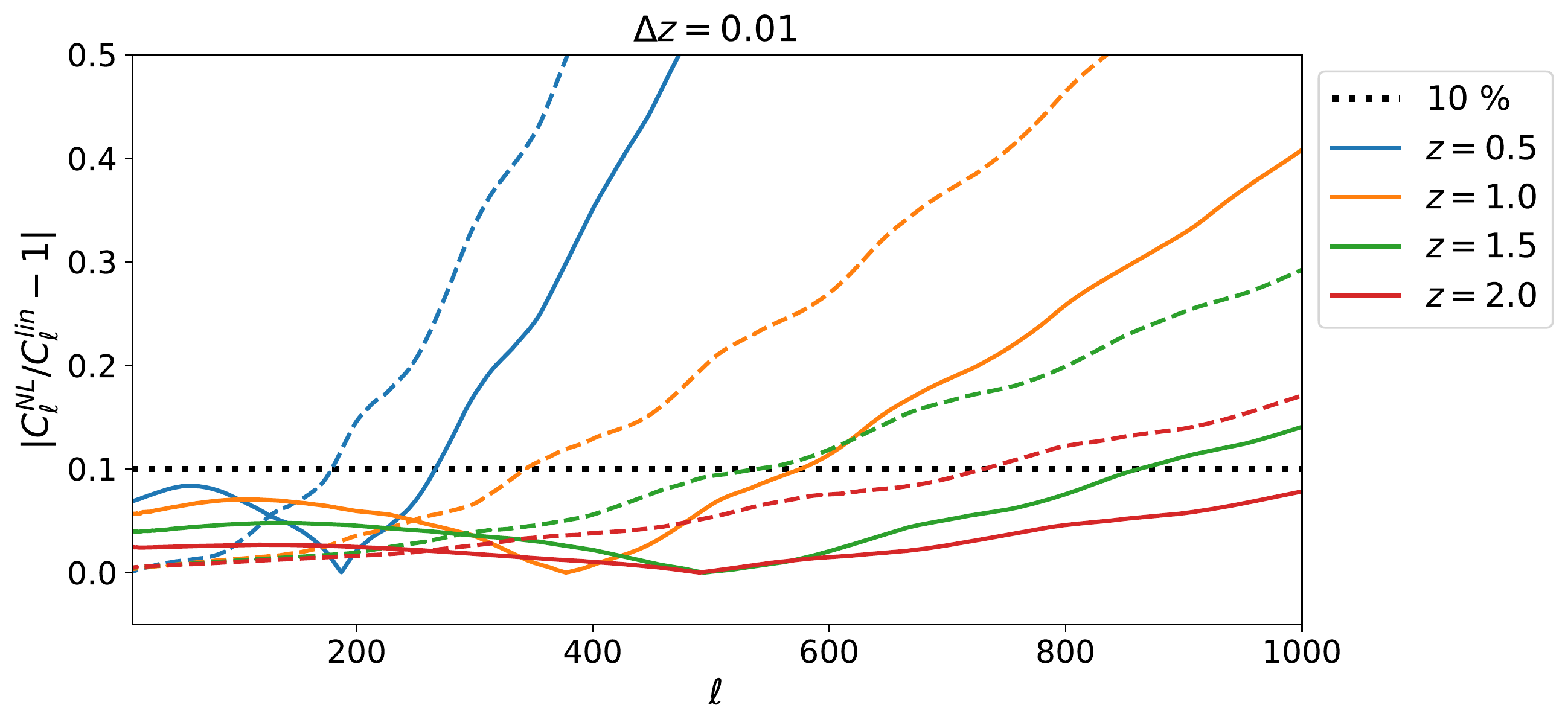}
    \includegraphics[scale=0.5]{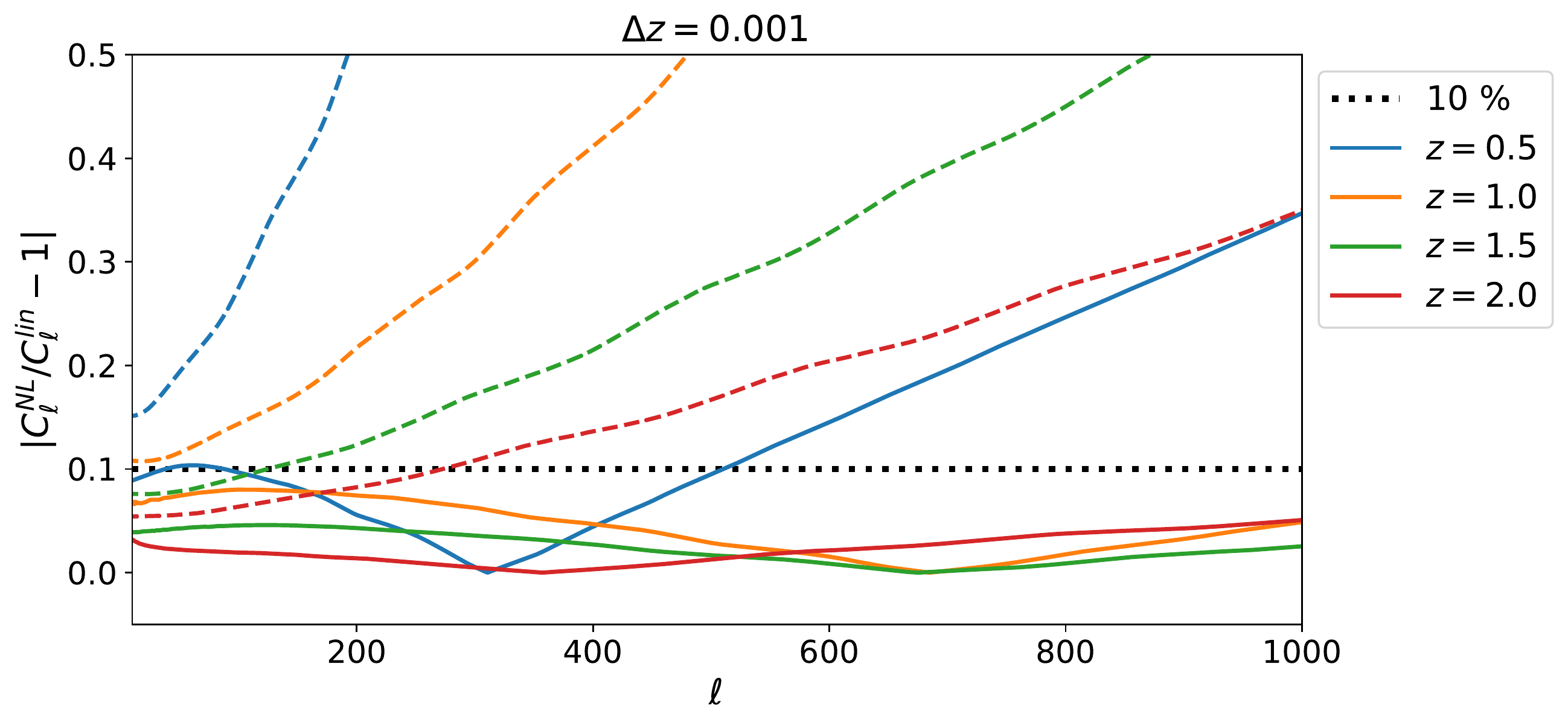}
    \caption{Relative difference between $C_\ell^{\rm NL}$ and $C_\ell^{\rm lin}$ for the TNS model for different redshifts. The dashed lines represent the density only (neglecting any RSD terms in the power spectrum) , and the solid ones represent density plus RSD. The horizontal dotted black line is the $10 \%$ line.
    \label{fig:reldiff_TNS}}
\end{figure}

\begin{figure}[htbp]
    \centering
     \includegraphics[scale=0.47]{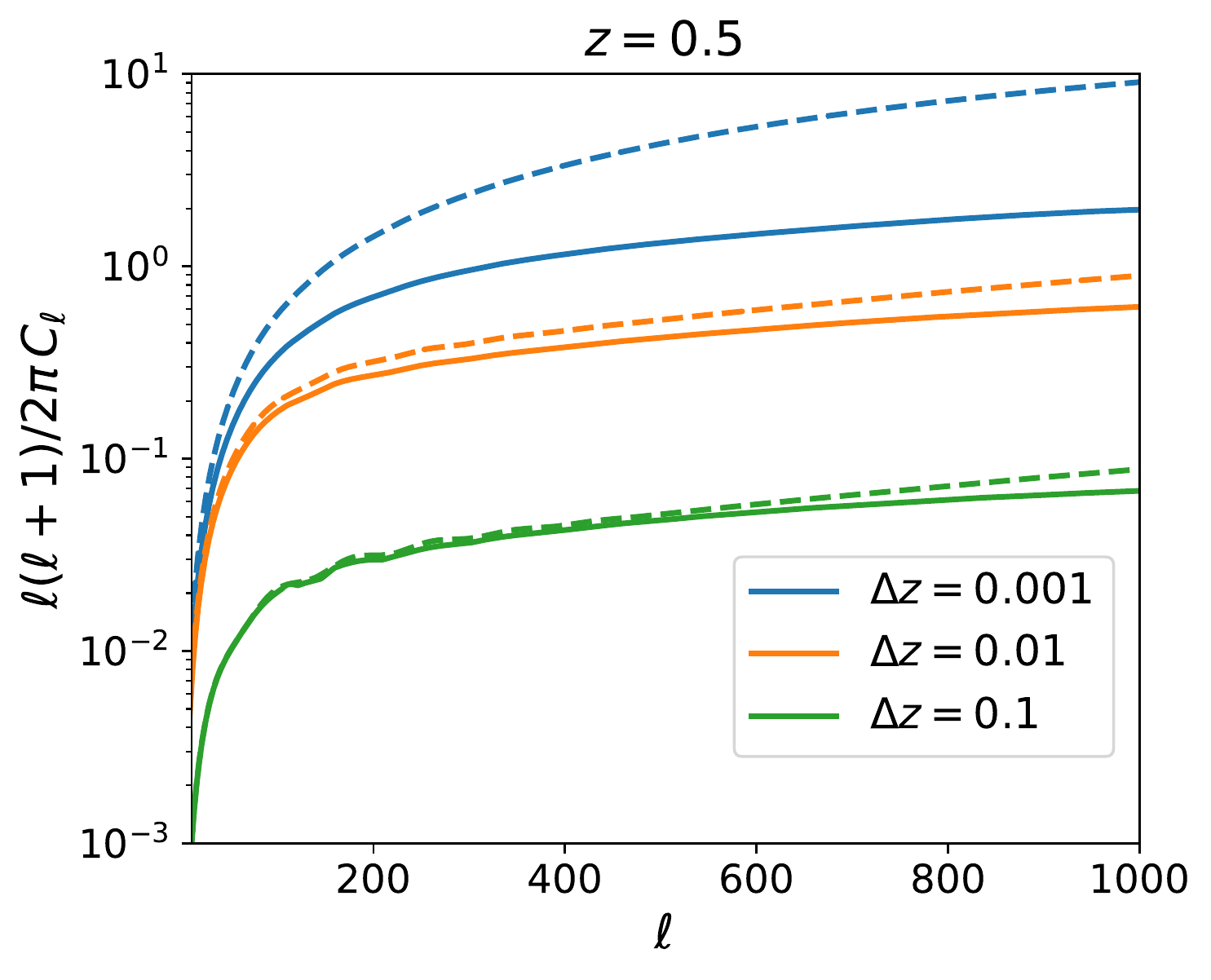}
     \includegraphics[scale=0.47]{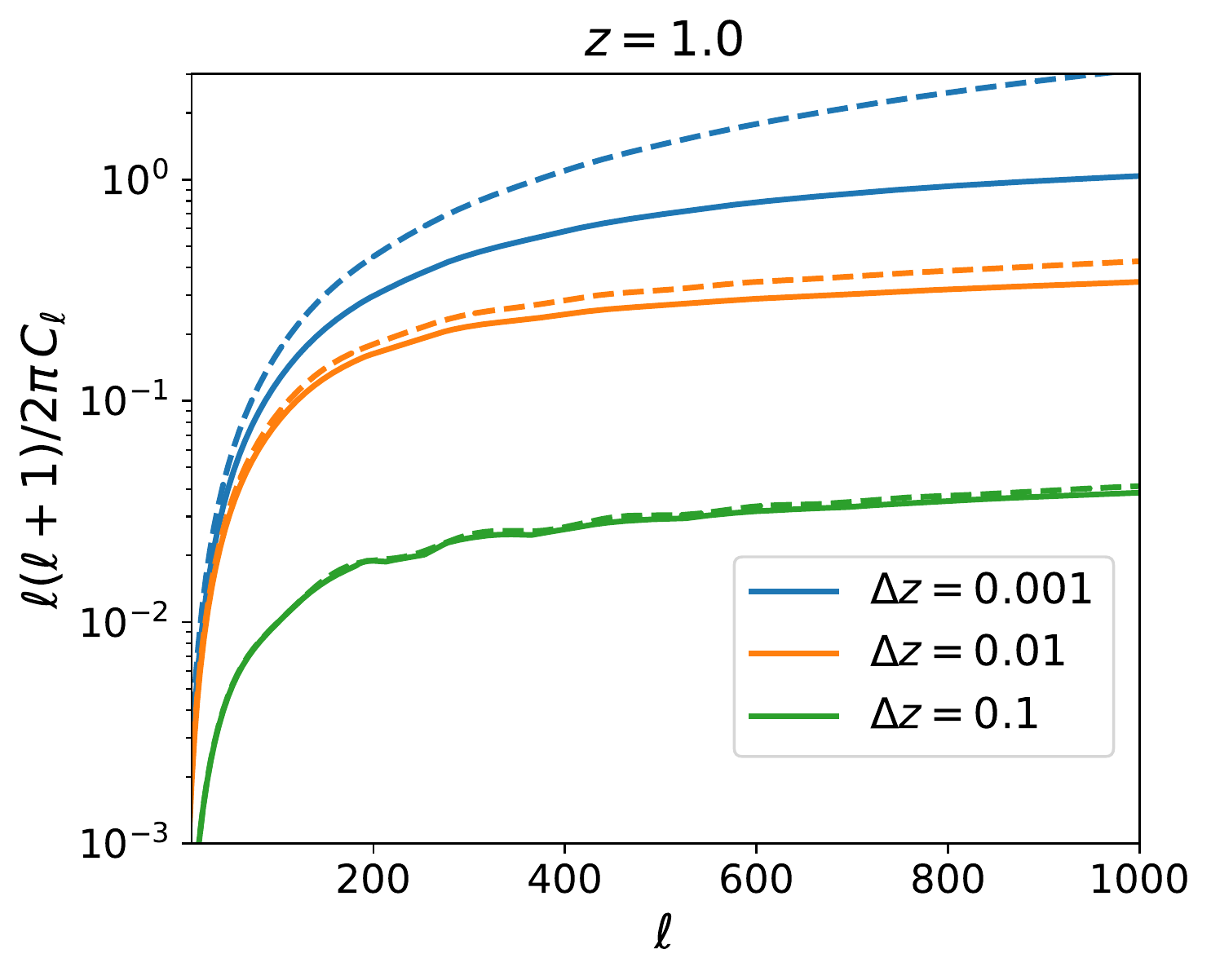}\\
     \includegraphics[scale=0.47]{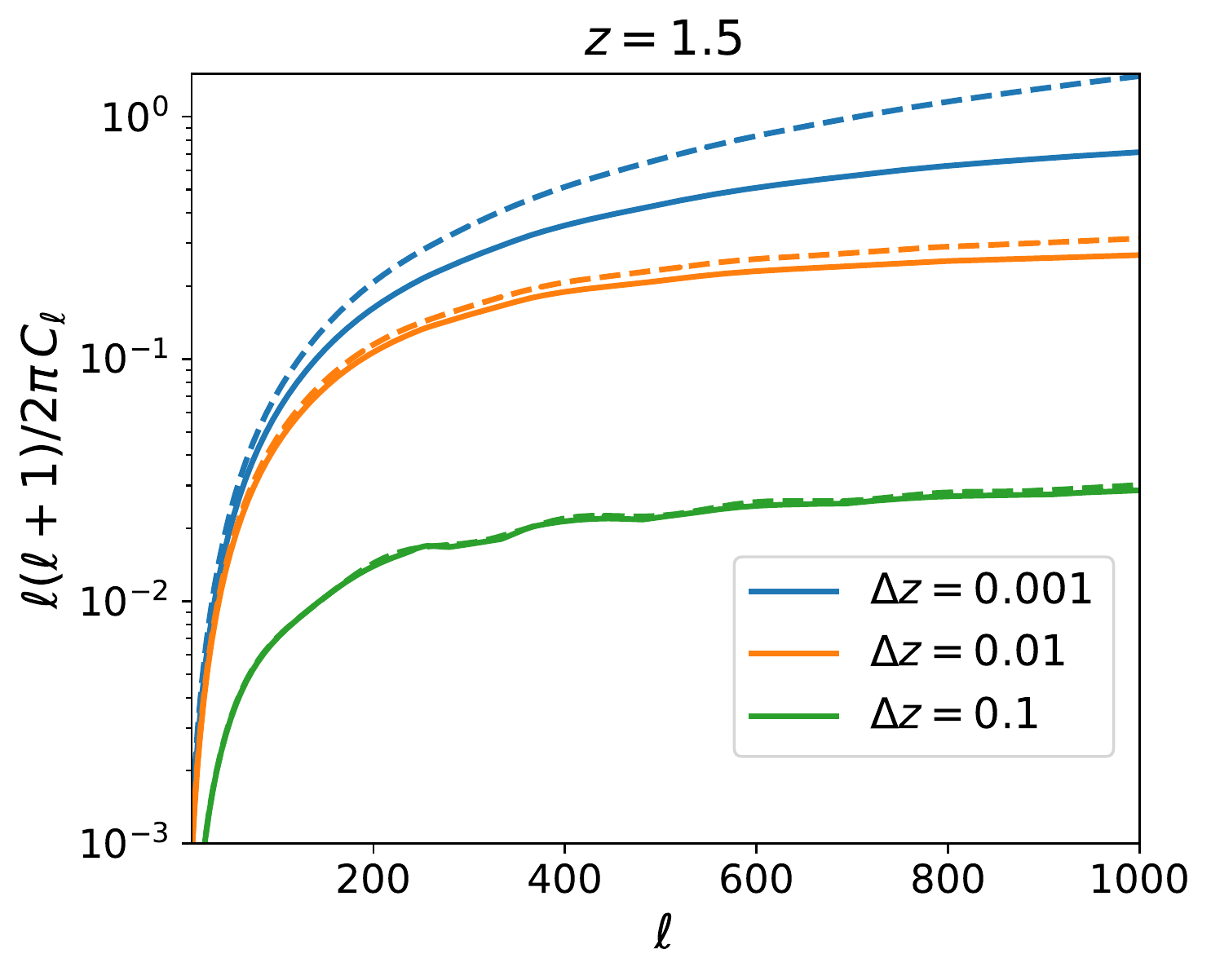}
     \includegraphics[scale=0.47]{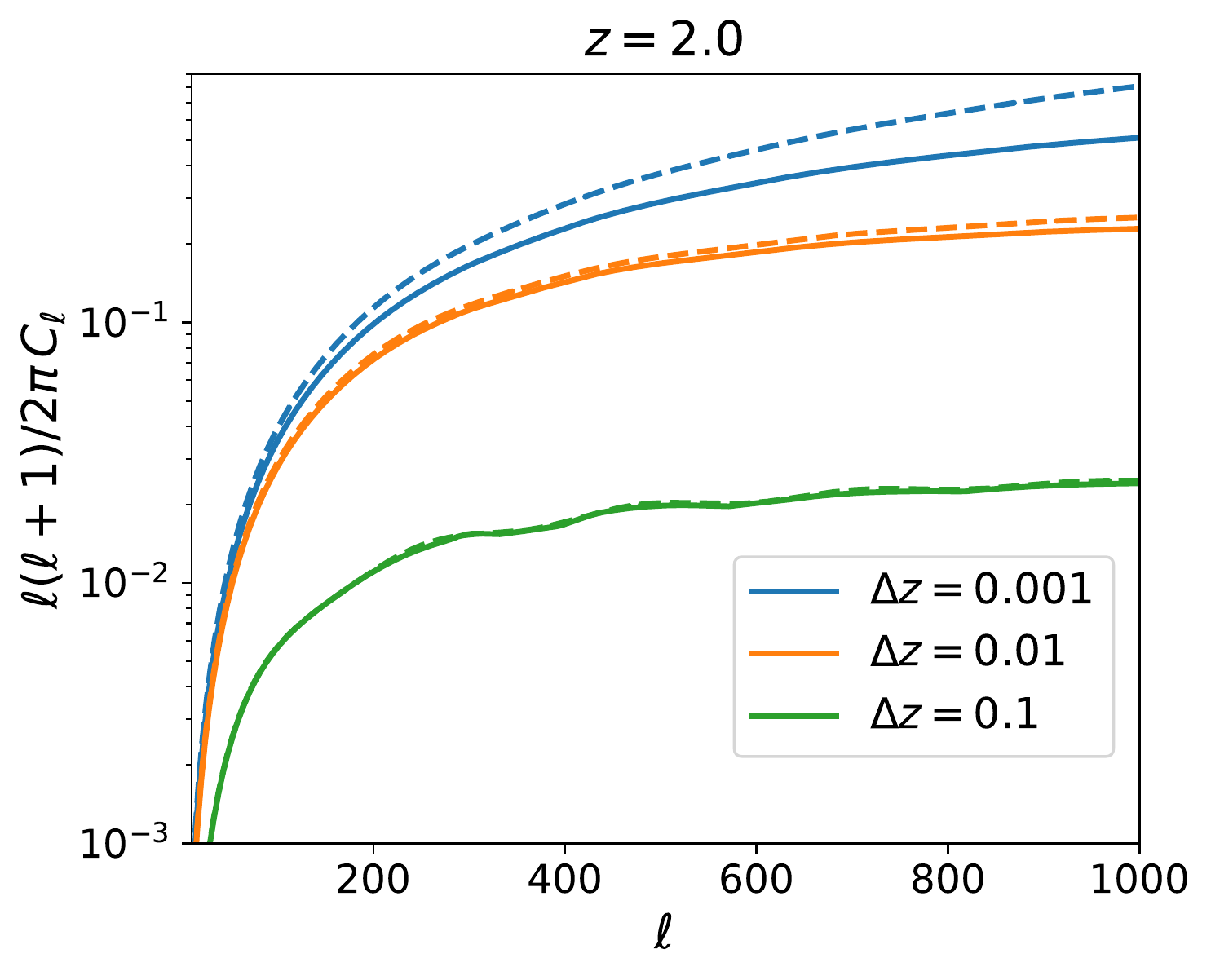}
    \caption{Comparison between CAMB halofit (dashed) and the COLA simulations (solid) in case of redshifts $z=0.5,\,1.0,\,1.5,\,2.0$ for different redshift bin widths.}
    \label{fig:CAMB_simul}
\end{figure}

\begin{figure}[htbp]
    \centering
     \includegraphics[scale=0.47]{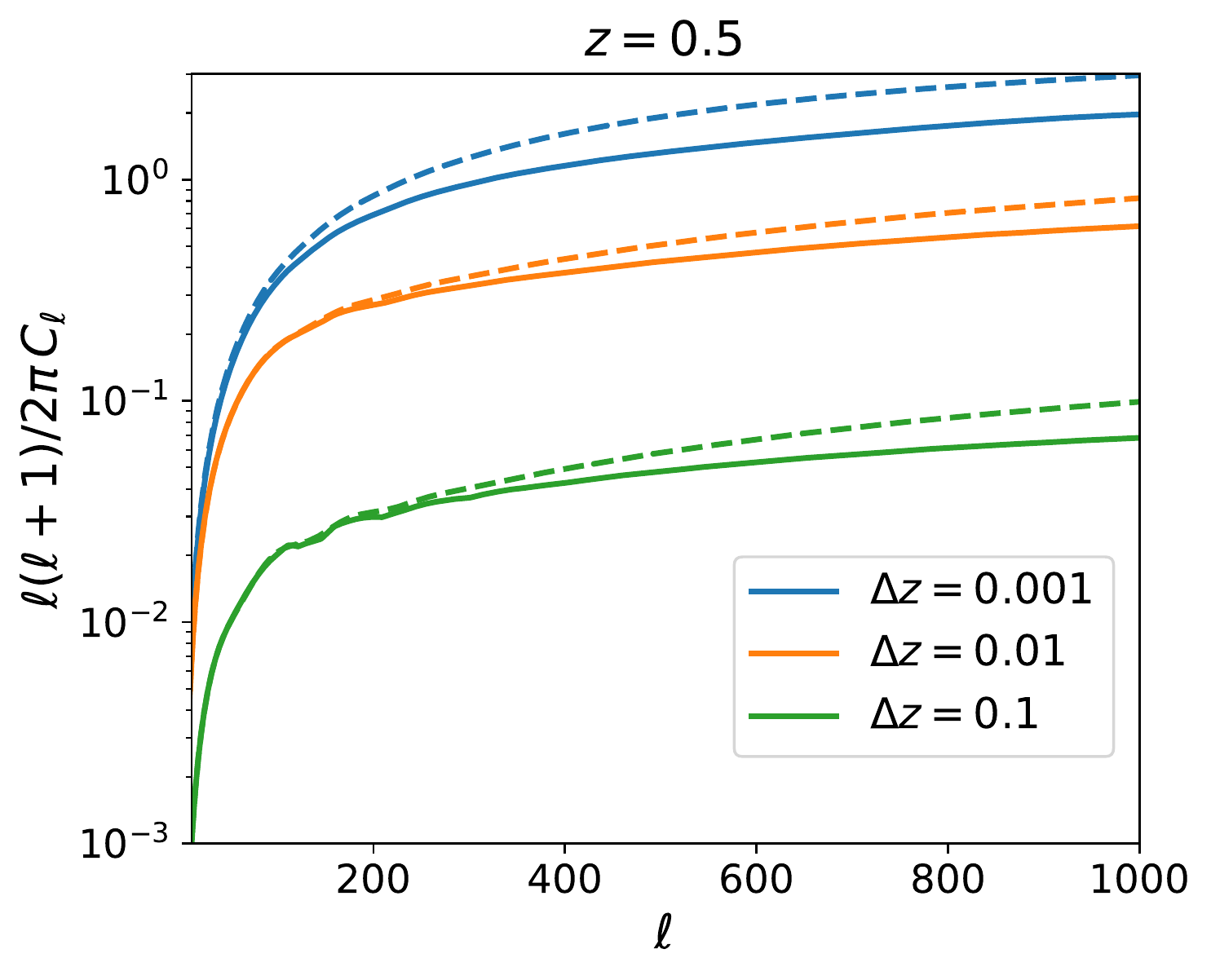}
     \includegraphics[scale=0.47]{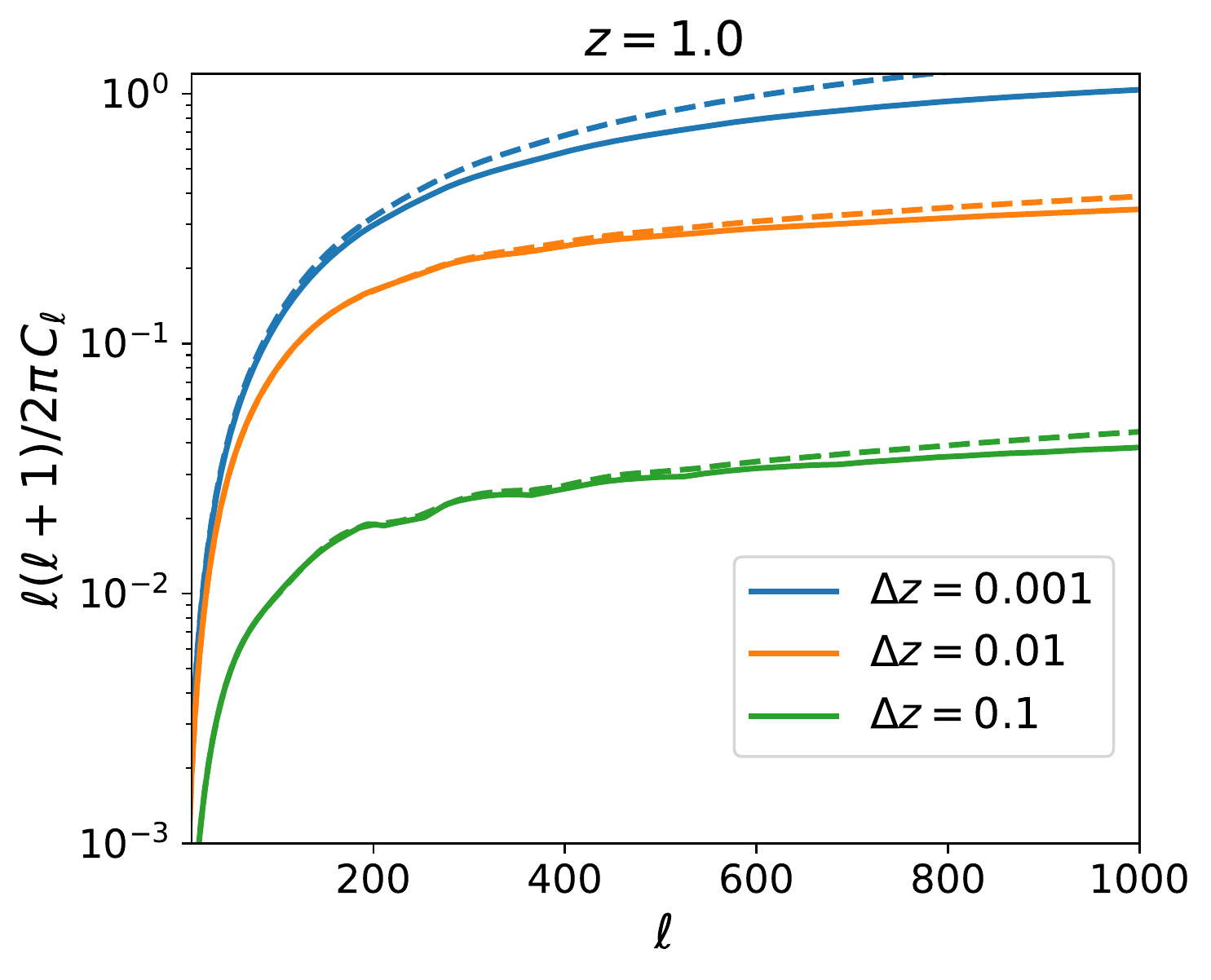}\\
     \includegraphics[scale=0.47]{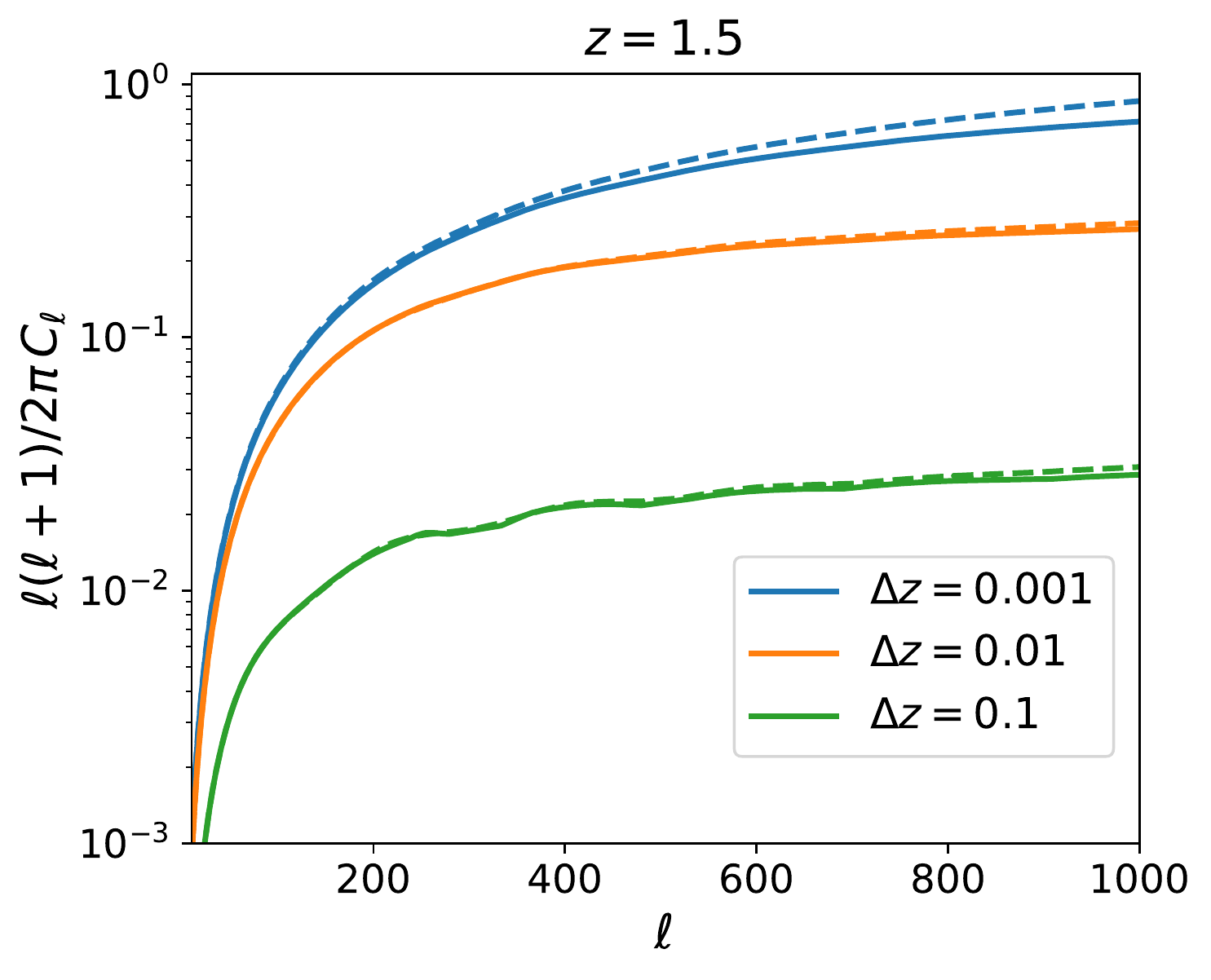}
     \includegraphics[scale=0.47]{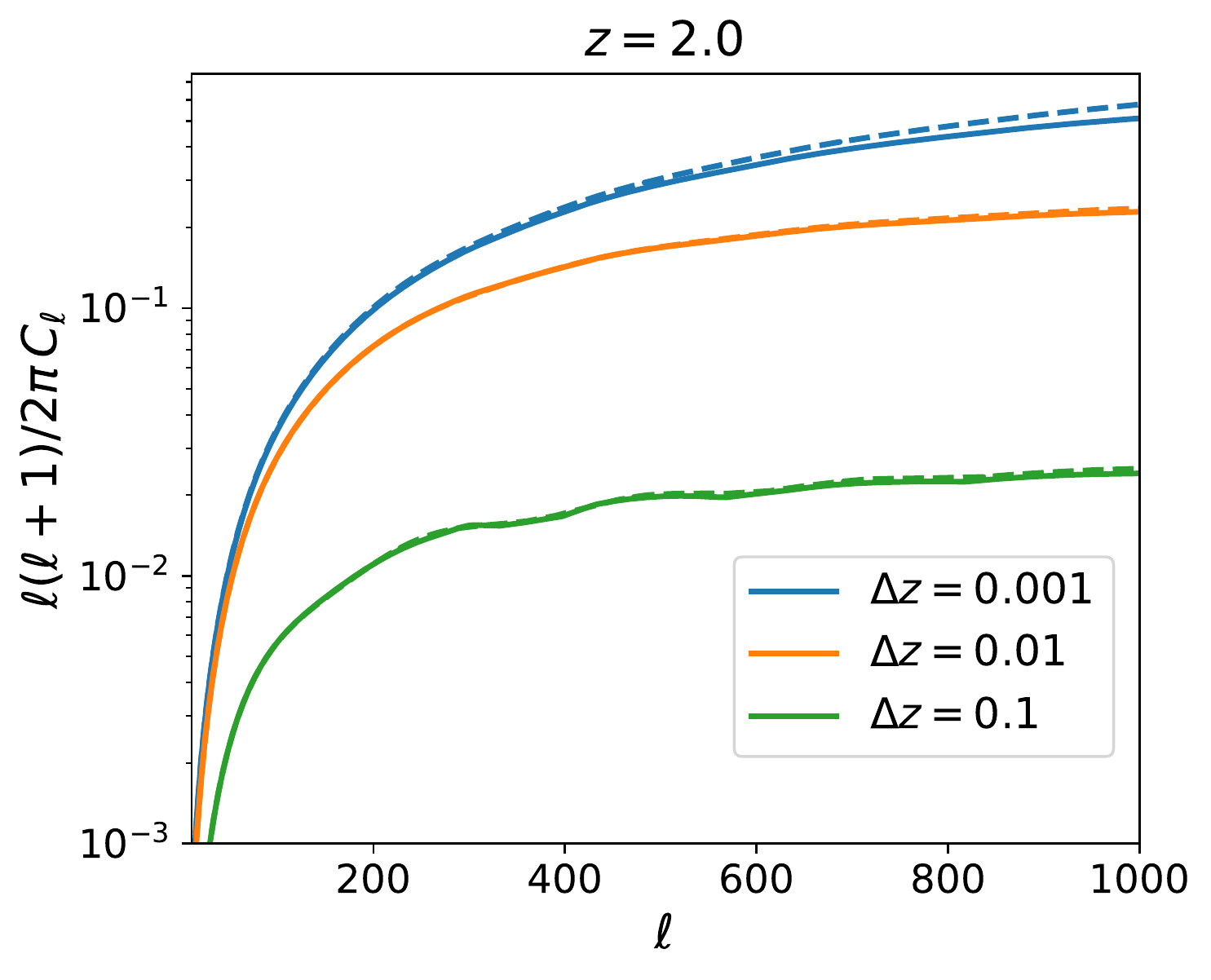}
    \caption{Comparison between TNS (dashed) and simulations (solid) in case of redshifts $z=0.5,\,1.0,\,1.5,\,2.0$ for different redshift bin widths.}
    \label{fig:TNS_simul}
\end{figure}

In Fig.~\ref{fig:reldiff_TNS} we compare linear and non-linear angular spectra for different redshifts using the TNS model for the non-linear case. As we explain later in Fig.~\ref{fig:CAMB_simul} and~\ref{fig:TNS_simul}, for the smaller redshift bins which are sensitive to RSD, TNS follows the simulation results better than CAMB halofit. For $\De z=0.1$, the higher the redshift the higher the value of $\ell$ below which our model deviates by less than 10\%  (black dotted line) from the linear result. Furthermore, redshift space distortions are not very  visible in $\ell$ space for $\Delta z=0.1$ (see top panel of Fig.~\ref{fig:reldiff_TNS}). 

For  $\Delta z=0.01$ and $\Delta z = 0.001$ (middle and low panels of Fig.~\ref{fig:reldiff_TNS}), RSD's are very prominent but now, even for very small $\ell$, the linear approximation is not sufficient any more. This is due to the fact that a very precise redshift resolution in the spectrum requires a very small radial mode, hence is sensitive to very high values of $k_\parallel$ which are affected by non-linearities.
Physically this just means that we are sensitive to non-linearities if we want high resolution in any direction, radial or transversal. This shift of the non-linear scale to lower $\ell$'s for narrow redshift bins is also visible in the lower panels of Fig.\ \ref{fig:Cell_compare}.

This is a very important result of the present paper: \emph{if we want to resolve RSD in the angular power spectrum we must have sufficiently precise redshift measurements, in which case the $C_\ell$'s are sensitive to non-linearities in the radial power spectrum.} 

At the highest redshift, $z=2$ and for the most narrow redshift bin, $\De z=0.001$ this yields simply a nearly constant offset from the linear results by about 5\%.  For lower redshifts and/or larger bin widths the difference from the linear result grows with $\ell$ as one naively expects. It is also interesting to note that for the smallest bin width (bottom panel of Fig.~\ref{fig:reldiff_TNS}), the deviation never exceeds 10\% for $z\geq 1$ or $z=0.5$ and $\ell\lesssim 500$. This can be understood by noting that RSD's which are most significant for the smallest bin width damp the non-linearities.

\begin{figure}
    \centering
    \includegraphics[scale=0.4]{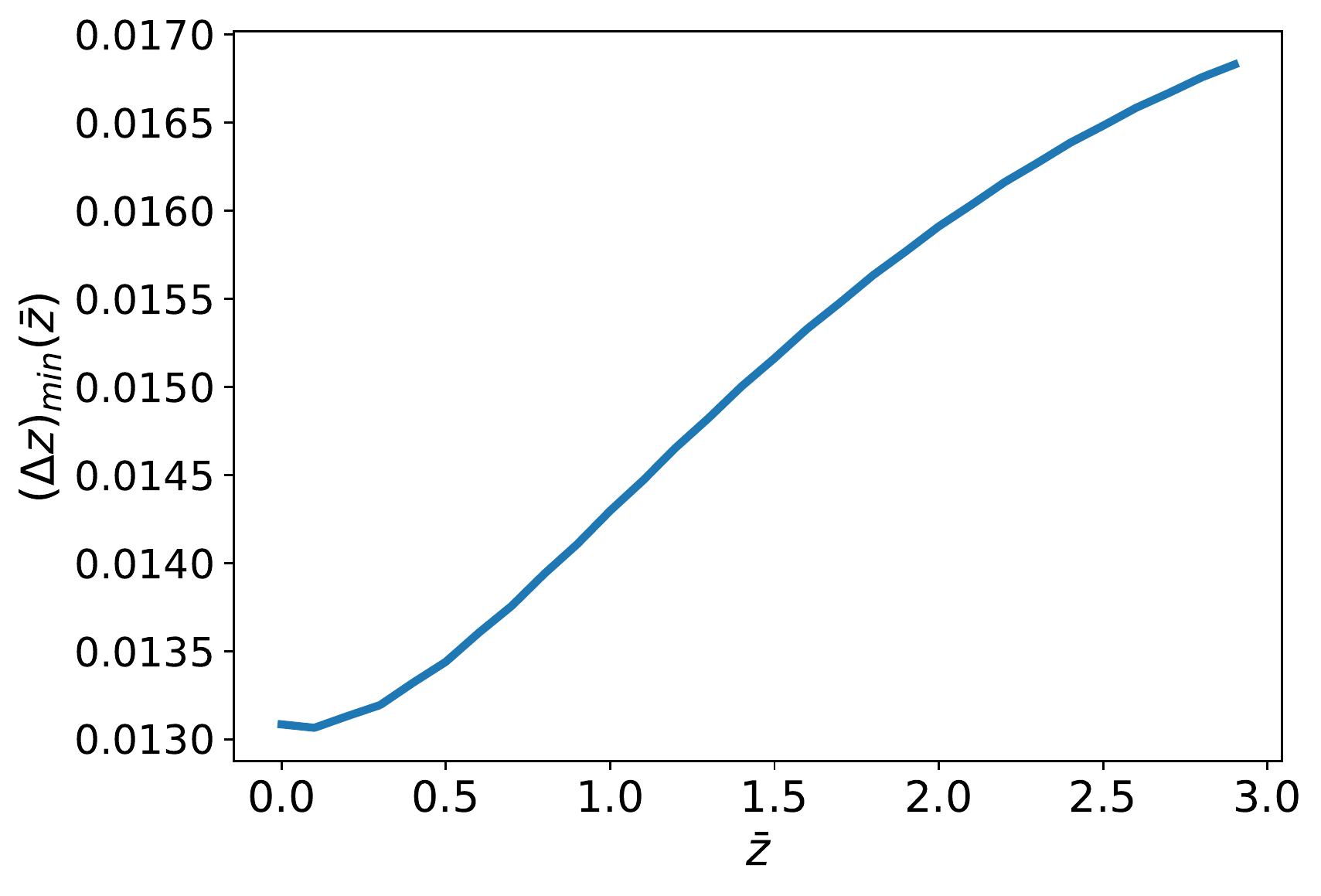}
    \includegraphics[scale=0.4]{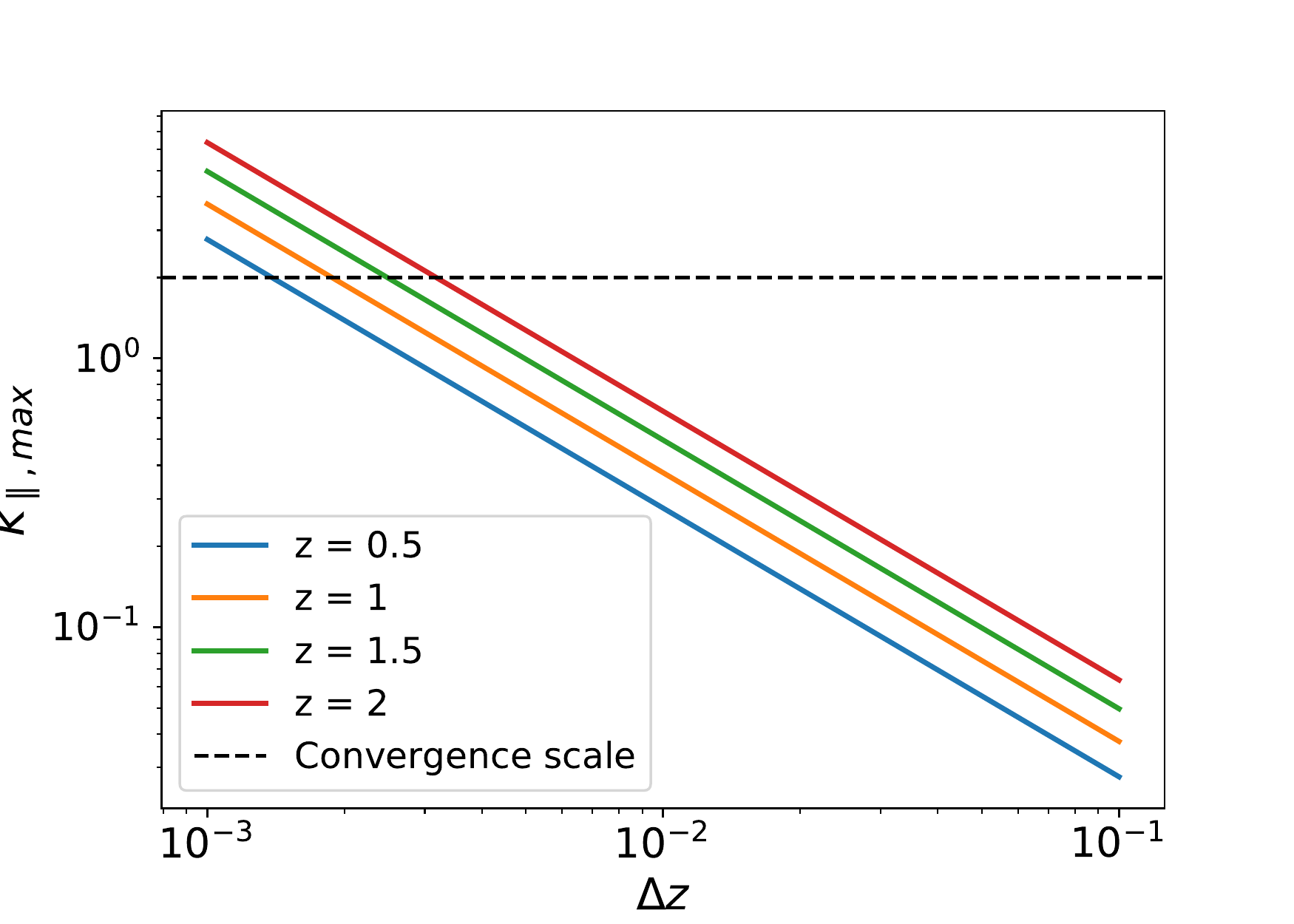}
    \caption{Left: $(\Delta z)_{\rm min}$ defined in Eq.~(\ref{eq:Deltaz_min}) as a function of $\bar{z}$. For redshift bins narrower  than $(\Delta z)_{\rm min}$ we expect radial non-linearities to affect the $C_\ell$'s also for low values of $\ell$. Right: The radial scale $k_{\parallel,\max}$ for which the integrand of the $C_\ell$  integral is damped by the Bessel function, as function of bin width $\Delta z$, for different redshifts. We also show as a dashed line the convergence scale of the integral in the case $\Delta z \rightarrow 0$.}
    \label{fig:Deltaz_min}
\end{figure}

Mathematically, the fact that non-linearities at small $\De z$ enter already at low $\ell$ can be understood very nicely from our flat sky approximation. Convolving Eq.~\eqref{eq:flat-sky-Ruth} with a tophat window function of width $\De z$ we find for a mean redshift denoted by $\bar z$
\be \label{eq:flatsky_bin}
C_\ell(\bar z,\De z)= \frac{1}{\pi \chi^2}\int dk_{\parallel}j_0^2\left(\frac{k_{\parallel}\De z}{2H(\bar z)}\right)P\left(k_{\parallel},\frac{\ell}{\chi}\right) \,.
\ee
Here the spherical Bessel function, $j^2_0(k_{\parallel}\De z/2H)$ acts as a `low pass filter' which filters out modes with $k_{\parallel}\gg 2H(z)/\De z$. For very small $\De z$ the integral therefore extends to high values of $k=\sqrt{k_{\parallel}^2 +(\ell/\chi)^2}$ for any $\ell$, and these modes can become large and non-linear. In this case non-linearities affect the result even at the lowest $\ell$ values. In other words, for linear perturbation theory to apply it is not sufficient that the relevant transverse modes, $k_{\perp} =\ell/\chi(z)$ are well in the linear regime, but also the relevant radial modes, $k_{\parallel}\leq 2H(z)/\De z$ must be in the linear regime. A crude approximation yields
\be \label{eq:Deltaz_min}
k_{\parallel,\max} \simeq \frac{2\pi H(\bar z)}{\De z} <k_{\rm NL}(\bar z) \quad \mbox{ or }\quad 
\De z \gtrsim (\De z)_{\min}= \frac{2\pi H(\bar z)}{k_{\rm NL}(\bar z)} \,.
\ee
We show $(\De z)_{\min}$ as a function of $\bar z$ in Fig.\ \ref{fig:Deltaz_min}. The critical width is therefore of the order of $\Delta z \approx 0.01$ to $0.02$, for narrower redshift bins (higher redshift resolution) we have to expect that (radial) non-linearities affect the $C_\ell$ for all values of $\ell$, not only for $\ell > \ell_{\rm NL}$. 

The radial cutoff scale $k_{\parallel,\max}$, also shown in Fig.\ \ref{fig:Deltaz_min}, lies well below the non-linear scale for $\Delta z = 0.1$, while for $\Delta z = 0.01$ it is in the range of $k \approx 0.3h$/Mpc to $0.8h/$Mpc, depending on redshift, already in the non-linear regime. For narrow redshift bins, $\Delta z = 0.001$, it becomes larger than the `absolute' convergence scale of $k\approx 2 h/$Mpc, for which the $C_\ell$ integral (\ref{eq:flatsky_bin}) converges without any damping from the Bessel function, i.e.\ also for $\Delta z \rightarrow 0$ (except for very high $\ell$ where the effective starting value of the integration, $\ell/\chi$, is pushed to higher $k$).

\begin{figure}[t]
    \centering
    \includegraphics[scale=0.5]{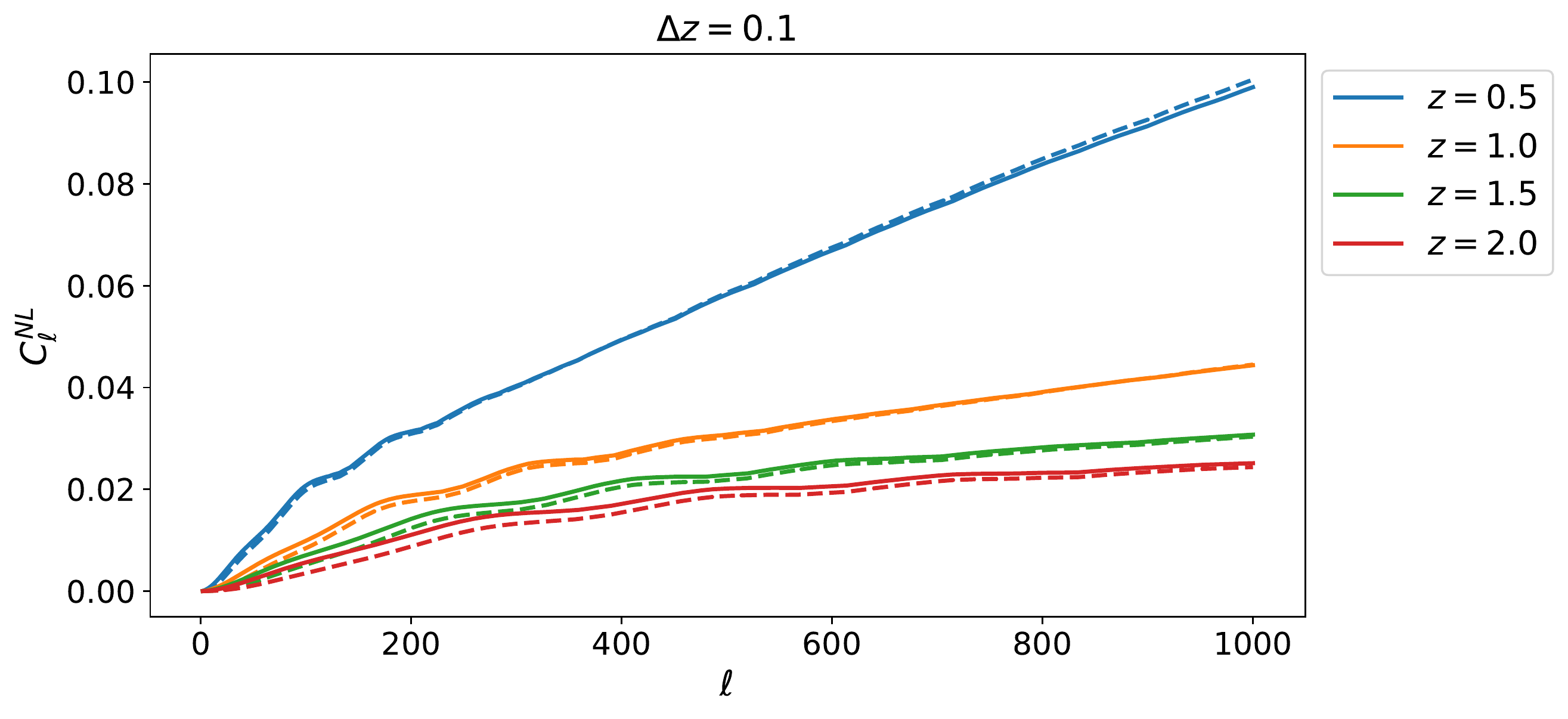}
    \includegraphics[scale=0.5]{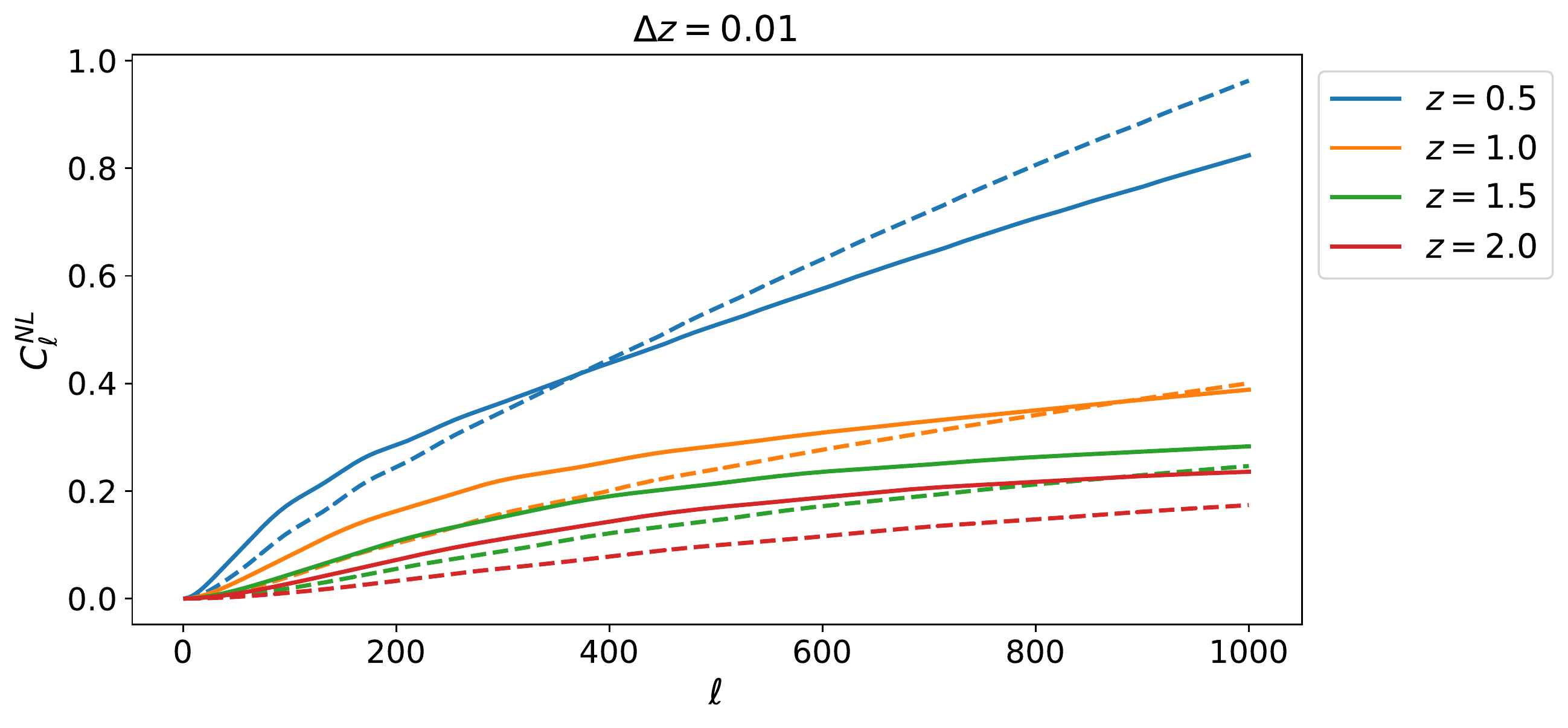}
    \includegraphics[scale=0.5]{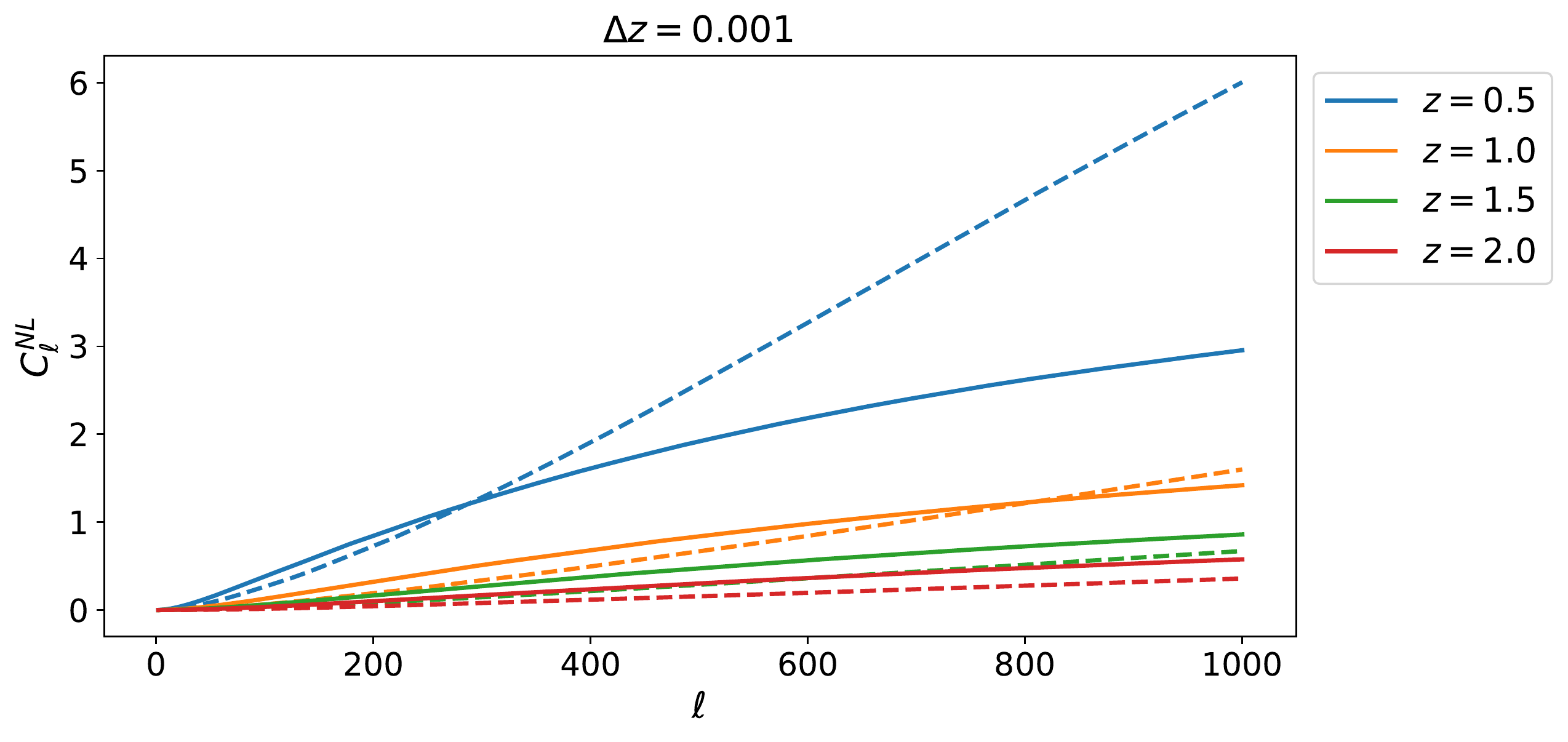}
    \caption{The comparison between $C^{\rm NL}_{\ell}$ for density and density + RSD, for different mean redshifts and with width $\Delta z=0.1,\,0.01,\,0.001$ from top to bottom. $NL$ stands for non-linear, and in this plot we show the case of TNS model. The dashed lines contain density only while in the the solid lines both, density and RSD are included.}
    \label{fig:TNS_dens_densrsd}
\end{figure}

\begin{figure}[htbp]
    \centering
    \includegraphics[scale=0.5]{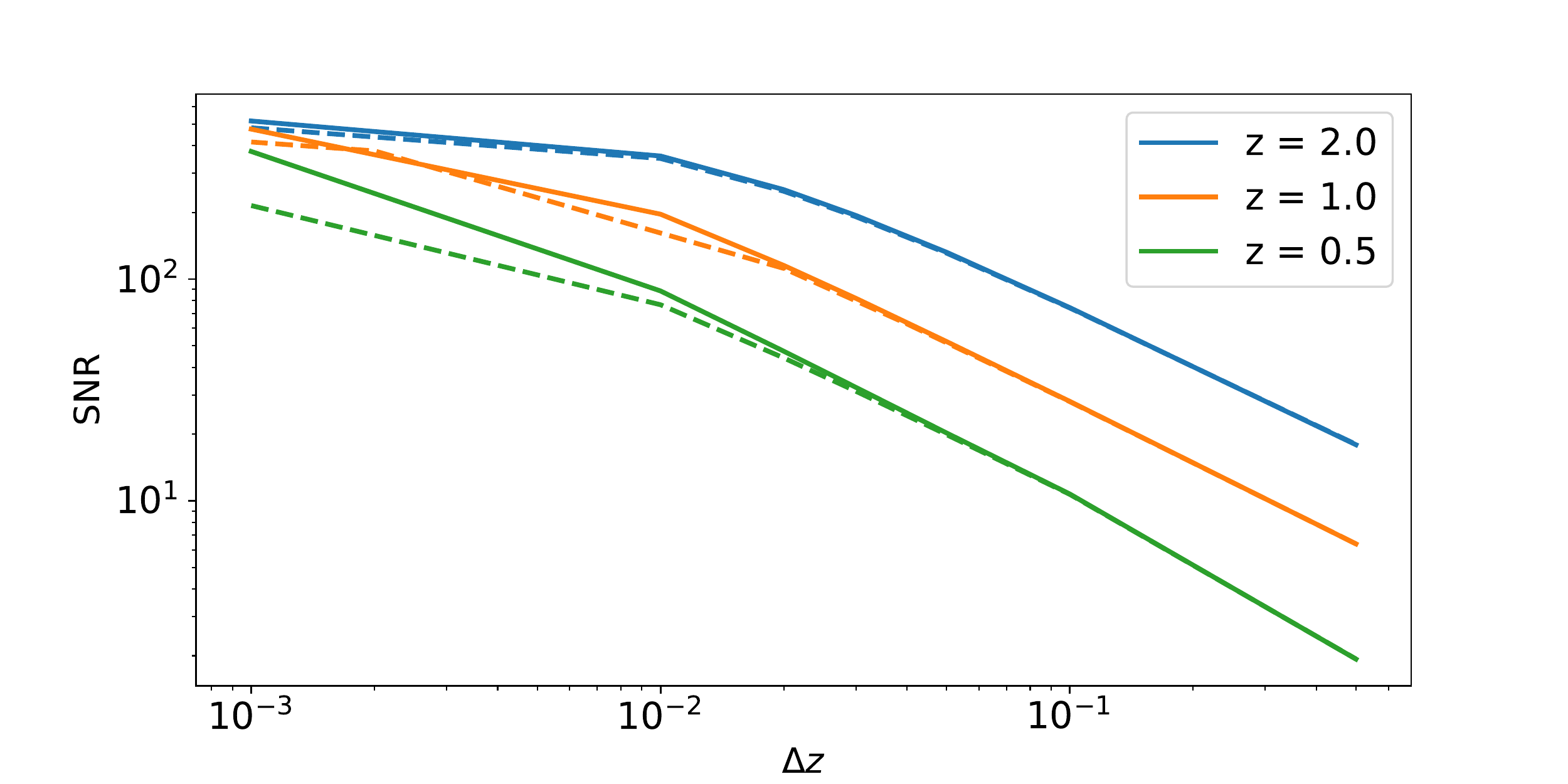}
    \caption{Signal-to-noise ratio for detecting RSD as a function of $\Delta z$ for different redshifts. Dashed lines represent SNR using linear $C_\ell$'s and solid lines represent SNR using non-linear $C_\ell$'s. For $z=1$ and $z=2$ we sum up to $\ell_{\rm max}=1000$ and for $z=0.5$, we sum up to $\ell_{\rm max}= 600$ which is $2\ell_{\rm NL}$. \label{fig:SNR}}
\end{figure}

To illustrate clearly the relevance of RSD's we show  the difference between the density only (dashed) and density plus RSD (solid) in the non-linear predictions in Fig.~\ref{fig:TNS_dens_densrsd} using the TNS approximation. For the widest redshift bin, $\De z=0.1$, redshift space distortions are not very relevant. For small $\De z$, however, they significantly reduce the $C_\ell$ spectrum at high $\ell$. As for the power spectrum, on linear scales RSD enhances the power spectrum via the Kaiser effect while on non-linear scales it reduces it due to the velocity overshoot which damps the density power spectrum in redshift space. The crossover between the dashed and the solid line roughly corresponds to the non-linearity scale at a given redshift. This explains also why the crossover location is nearly independent of the bin width $\De z$.
\\
To quantify the importance of RSD terms, we have performed a simple Fisher forecast for several bin widths. We modeled the RSD's with the Kaiser formula applied to halofit, and only kept the cosmic variance contribution to the noise (neglecting survey-dependent contributions like shot noise and sky fraction). More details about the Fisher analysis are given in Appendix~\ref{app:Fisher}. The signal-to-noise ratio (SNR) is shown in Fig.\ \ref{fig:SNR} as a function of redshift bin width, $\Delta z$, for three different redshifts. As we see, the RSD signal drops by an order of magnitude when going from $\Delta z = 0.01$ to $\Delta z = 0.1$, highlighting the importance of using narrow redshift bins for measuring RSD. The RSD signal is however still detectable even for wide bins, and it should therefore be included in the $C_\ell$ also for $\Delta z = 0.1$.

In Fig.~\ref{fig:CAMB_simul} we compare also the result of halofit from CAMB (dashed)  with the one from the COLA simulations (solid) for density plus RSD angular power spectra. For the bin width $\De z=0.1$ we only have a slight overshoot of the spectrum at $\ell>400$ for $z=0.5$, all other spectra are in good agreement. However, for small bin widths $\De z\leq 0.01$, and especially for $\De z=0.001$, the insufficient treatment of the RSD in the halofit model where they are taken into account simply by the linear Kaiser formula, leads to a significant spurious amplification of the power spectrum already at low values of $\ell$. This overshoot is more significant at lower redshifts, where nonlinearities are more relevant, but it is already visible at $z=2$.

In Fig.~\ref{fig:TNS_simul} we compare the results using the TNS approximation (dashed) with the one from the COLA simulations (solid) for density plus RSD angular power spectra. Clearly, the TNS approximation handles redshift space distortions much better that halofit and the spurious excess is reduced and no longer visible for $z=1.5$ and $2.0$. However, we have checked that the relative difference between the simulation result and TNS is larger than cosmic variance for $z=2$ and $\ell>200$ as well as $z\leq 1.5$ and $\ell=100$ for narrow bins, $\De z=0.001$. It becomes larger than cosmic variance at $\ell=$ 700, 350 and 200 for $\bar z=$ 1.5, 1 and 0.5 respectively for both bin widths, $\De z=0.01$ and $\De z=0.1$. This indicates again that for slim redshift bins, $\De z=0.001$, non-linearities are relevant already at low $\ell$ (where however cosmic variance is very large).
For wide redshift bins, $\De z=0.1$, where RSD are not very relevant, the halofit model from CAMB is actually a better approximation than TNS. We have already seen this in Fig.~\ref{fig:Cell_compare}.


\section{Discussion and conclusion}\label{s:con}
In this paper we have compared different perturbation theory-based schemes to treat non-linearities in the angular power spectrum. We also compare these predictions with COLA simulations. These simulations are percent level accurate within $k\lesssim 1h/$Mpc when compared to full N-body measurements of the matter power spectrum \cite{Winther:2017jof}. At the level of the redshift space multipoles, the simulations still provide fair accuracy up to $k\sim 0.3h/$Mpc \cite{Bose:2019psj}. All other models considered here are less accurate and so we use the COLA simulations as our benchmark in accuracy. 

In particular, we compare standard perturbation theory (SPT), Lagrangian perturbation theory (LPT), the effective field theory of large scale structure (EFT) and the TNS model. For these approaches, we provide two bases of comparisons. The first is at the level of the two dimensional redshift space power spectrum $P(k,\mu)$. At this level we make the following conclusions:
\begin{itemize}
    \item 
    The TNS model offers the best modelling of the RSD anisotropy, being comparable to the COLA measurements into the quasi linear regime, $k\lesssim 0.2h/{\rm Mpc}$, at $z=0.5$, and over a wide range of $\mu$.
    \item 
    LPT and SPT do the worst depending on the value of $\mu$. At $\mu=1$ where the Fingers of God effect dominates, LPT out-performs SPT while the two do comparably well at $\mu=0$.
    \item
    Non-linear RSD modelling is essential in modelling the 2D power spectrum. This has been checked by comparing the Kaiser formula combined with the halofit non-linear matter power spectrum to the COLA measurements. Despite being a good approximation at $\mu \sim 0$, it performs the worst out of all models at $\mu \sim 1$ and $k \lesssim 0.1 h/$Mpc.
\end{itemize}

The second basis for comparison is at the level of the angular power spectrum. We note that this quantity is more directly related to our observations. Our main conclusions are the following: 

\begin{itemize}
    \item 
    We find that the flat sky approximation is valid at percent level accuracy and so adopt this for all our comparisons of $C_\ell$.
    \item 
    At large bin widths ($\Delta z \sim 0.1$), RSD is much less important and the main contributor to non-linear information is within the matter power spectrum. At this bin width the most accurate prescription is halofit combined with the Kaiser factor, being accurate to within a few percent of the COLA measurement up to $\ell \lesssim 400$ at $z=0.5$.
    \item 
    Small bin widths greatly enhance the impact of non-linear RSD. Because of this, the TNS model out-performs all other models for $\Delta z = 0.01$ and $\Delta z =0.001$. Despite this, it is still a poor approximation, being accurate to within a few percent for $\ell \lesssim 150$ at $z=0.5$. 
    \item
    For small bin widths, non-linear RSD information becomes important at very small $\ell$, with very large non-linear effects ($\sim 10\%$) being found at $\ell \leq 50$ at $z=0.5$ for the TNS model.
    \item
    For large bin widths, the effect of lensing cannot be ignored, and at $z=0.5$ with $\Delta z = 0.1$, it is already equal in magnitude to the RSD signal at $\ell \leq 150$ (see Appendix~\ref{app:lensing}).
\end{itemize}

In conclusion, at the level of the angular power spectrum, it becomes very difficult to disentangle non-linearities and various contributions to the signal. In particular, at low redshift, non-linear RSD can play a large role at $\ell \leq 150$ for small bin width choices while for large bin widths lensing begins to dominate the signal. At high redshift ($z > 1$)  non-linear RSD is better controlled but lensing becomes more important for large bin widths.  At $z=1$ lensing is sub-dominant to RSD up to $\ell \lesssim 500$ for small and large bin width choices. The TNS model offers a relatively good prescription to model the non-linear effects of RSD in the angular power spectrum, but is still very limited, especially at low redshift where non-linearities are enhanced. 

While this can be circumvented by only considering spectra for `linear' $\ell$ for wide redshift bins, this becomes impossible for narrow redshift bins. In fact, for $\Delta z \lesssim 0.001$, where we need to accurately model the non-linear spectrum to high $k$ for all values of $\ell$, no prescription is currently accurate enough. It appears therefore that at least for now the angular spectrum is less well suited to measure RSD's than the correlation function.

We have found that while the TNS approximation is the only one with a reasonably good treatment of velocities, it does not reproduce well the COLA angular power spectra for wide redshift bins, $\De z\geq 0.1$. For such wide-bin spectra, RSD's are not important and halofit, which gives the better fit to the density only power spectrum than TNS, is actually preferable. On the other hand, for slim redshift bins, $\De z\leq 0.01$ TNS is a much better approximation. For such bin widths, radial non-linearities are already relevant for very small $\ell$'s which renders halofit, or even more so the linear power spectrum, simply useless. On the other hand, on scales $\ell>\ell_{\rm NL}(z)$, where also the transverse wave number enters the non-linear regime, also the TNS approximation which models the pure matter density power spectrum becomes insufficient, especially at low redshift, $z\simeq 0.5$. 

From this work it is clear that we are still far away from modelling the angular power spectrum at 1\% precision over a reasonable range of $\ell$. But we now know better in which direction we have to make progress. We need to model the density power spectrum similar to halofit but then correct for non-linear RSD like in the TNS model. Especially, if we want to model the $C_\ell$'s in narrow redshift bins where they are sensitive to redshift space distortions. This is essential if we wish to safely extract very important cosmological information. We must make sure to model RSD very precisely, as they can enter the $C_\ell$'s at small $\ell\ll \ell_{\rm NL}(z)$ depending on the bin width.

\section*{Acknowledgements}
It is a pleasure to thank Joyce Byun, Marko Simonovic, Zvonimir Vlah, Alberto Salvio, Farbod Hassani and Antony Lewis for helpful discussions.  
We acknowledge financial support from the Swiss National Science Foundation. BB acknowledges support from the Swiss National Science Foundation (SNSF) Professorship grant No.170547.
\vspace{0.65cm}

\newpage
\appendix
\begin{center}
\center{\bf\large APPENDIX \vspace{0.5cm}}
\end{center}
\section{Derivation of the 1-loop terms\label{Appendix:A}}

Following the notations and conventions in \cite{Heavens:1998es} for Eq.~(\ref{eq:ptot_SPT}), we have:
\begin{eqnarray}\label{e:heavappend}
    P^s_{tot}(\textbf{k})&\equiv& P^s_{11}+P^s_{22}+P^s_{13}\\ \nonumber
    &=& (1+\beta\mu^2)^2b_1^2P_{11}(k)+2\int\frac{d^3\textbf{q}}{(2\pi)^3}P_{11}(q)P_{11}(|\textbf{k-q}|)[F_2^S(\textbf{q},\textbf{k-q})]^2\\ \nonumber
    &&\quad +6(1+\beta\mu^2)b_1 P_{11}(k)\int \frac{d^3\textbf{q}}{(2\pi)^3}P_{11}(q) F_3^S(\textbf{q},\textbf{-q},\textbf{k})
\end{eqnarray}
where $\beta\equiv f/b_1$ and $b_1$ denotes the linear bias. The symmetrised expression for $F_2^S(\textbf{k}_1,\textbf{k}_2)$ and the unsymmetrised one for $F_3(\textbf{k}_1,\textbf{k}_2,\textbf{k}_3)$ are shown in Eq.~(13) of \cite{Heavens:1998es}. We symmetrise $F_3(\textbf{k}_1,\textbf{k}_2,\textbf{k}_3)$ and find $F_3^S(\textbf{q},\textbf{-q},\textbf{k})$, neglecting higher order biases.\par
\par
\begin{figure}
    \centering
    \includegraphics[scale=0.45]{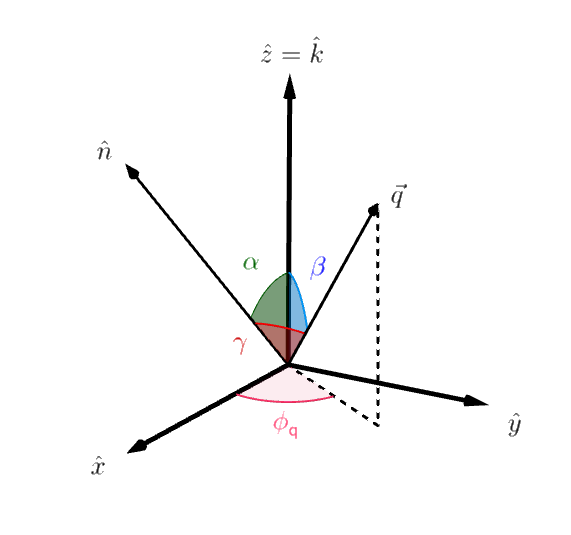}
    \caption{Vectors and angles involved in the calculations of $P_{13}$ and $P_{22}$: $\hat{\kvec}$ is the direction of the wave vector $\kvec$ of Eq.~(\ref{eq:ptot_SPT}) and $\qvec$ is also given in Eq.~(\ref{eq:ptot_SPT}).  $\mathbf{n}$ is the line-of-sight direction.}
    \label{fig:angles}
\end{figure}

 As can be seen in Eq.~(13) of \cite{Heavens:1998es}, the expressions for $F_2^S$ and $F_3^S$ are given in terms of $J_2^S$, $J_3^S$, $K_2^S$ and $K_3^S$, which can be computed from the general $n$th order expression as found in literature (see for example Eq.~(10a) and (10b) of \cite{Jain:1993jh}). While Eq.~(\ref{e:j2s}) and Eq.~(\ref{e:k2s}) given below are easily available in literature, for obtaining  Eq.~(\ref{e:j3s}) and Eq.~(\ref{e:k3s}), we have used the expression for $n=3$ and symmetrised it. However, our results did not match very accurately with the symmetrised expression obtained from Eq.\ (11) of \cite{Heavens:1998es}, and therefore we explicitly write them below in Eq.~(\ref{e:j3s}) and Eq.~(\ref{e:k3s}). We find these relations to be as follows:
\begin{eqnarray}
J_2^S(\textbf{q}_1,\textbf{q}_2)=\frac{5}{7}+\frac{1}{2}\frac{\textbf{q}_1.\textbf{q}_2}{q_1q_2}\left(\frac{q_1}{q_2}+\frac{q_2}{q_1}\right)+\frac{2}{7}\frac{(\textbf{q}_1.\textbf{q}_2)^2}{q_1^2q_2^2} \label{e:j2s}\\
K_2^S(\textbf{q}_1,\textbf{q}_2)=\frac{3}{7}+\frac{1}{2}\frac{\textbf{q}_1.\textbf{q}_2}{q_1q_2}\left(\frac{q_1}{q_2}+\frac{q_2}{q_1}\right)+\frac{4}{7}\frac{(\textbf{q}_1.\textbf{q}_2)^2}{q_1^2q_2^2}
\label{e:k2s}
\end{eqnarray}

\begin{align}
\label{e:j3s}
 J_3^S(\textbf{q}_1,\textbf{q}_2,\textbf{q}_3)&=\frac{1}{3} \text{Sym}\left[7\frac{\textbf{q}.\textbf{q}_1}{q_1^2}J_2^S(\textbf{q}_2,\textbf{q}_3)+\frac{q^2\textbf{q}_1.(\textbf{q}_2+\textbf{q}_3)}{q_1^2|\textbf{q}_2+\textbf{q}_3|2}K_2^S(\textbf{q}_2,\textbf{q}_3) \right. \\ \nonumber
  &\left. \quad +\left(7\frac{\textbf{q}.(\textbf{q}_1+\textbf{q}_2)}{|\textbf{q}_1+\textbf{q}_2|^2}+\frac{q^2(\textbf{q}_1+\textbf{q}_2).\textbf{q}_3}{|\textbf{q}_1+\textbf{q}_2|^2q_3^2}\right)K_2^S(\textbf{q}_1,\textbf{q}_2)
) \right]
\end{align}

\begin{align}
\label{e:k3s}
 K_3^S(\textbf{q}_1,\textbf{q}_2,\textbf{q}_3)&=\frac{1}{3}\text{Sym} \left[ \frac{\textbf{q}_1.\textbf{q}}{q_1^2}J_2^S(\textbf{q}_2,\textbf{q}_3)+\frac{q^2\textbf{q}_1.(\textbf{q}_2+\textbf{q}_3)}{q_1^2|\textbf{q}_2+\textbf{q}_3|2}K_2^S(\textbf{q}_2,\textbf{q}_3) \right. \\ \nonumber
 &\left. \quad +\left(\frac{\textbf{q}.(\textbf{q}_1+\textbf{q}_2)}{|\textbf{q}_1+\textbf{q}_2|^2}+\frac{q^2(\textbf{q}_1+\textbf{q}_2).\textbf{q}_3}{|\textbf{q}_1+\textbf{q}_2|^2q_3^2}\right)K_2^S(\textbf{q}_1,\textbf{q}_2)\right]
\end{align}
Here `Sym' indicates symmetrisation in $\textbf{q}_1$, $\textbf{q}_2$ and $\textbf{q}_3$.
One can replace $\textbf{q}_1$, $\textbf{q}_2$, $\textbf{q}_3$ and $\textbf{q}=\textbf{q}_1+\textbf{q}_2+\textbf{q}_3$ as required and effectively calculate these kernels.\\
The final expressions for $F_2^S(\textbf{q,k-q})$ and $F_3^S(\textbf{q},\textbf{-q},\textbf{k})$ along with subsequent calculations can be found in a Mathematica notebook whose link we will provide in an upcoming version.\\
In the kernels  $F_2^S$ and $F_3^S$ we encounter the scalar products $\hat\bq\cdot\bn \equiv  \mu_q = cos(\gamma)$, $\hat\bk\cdot\bn \equiv  \mu = cos (\alpha) $, and $ \hat{\bk} \cdot \bq \equiv x = cos(\beta)$. We also define $r=|\bq|/|\bk|$. We can write $\mu_q$ in terms of $\mu$, $x$, and $\phi_q$, where $\phi_q$ is the angle between the projection of $\bq$ and $\mathbf{n}$ onto the plane perpendicular to $\hat{\mathbf{k}}$ (see Fig.~\ref{fig:angles})
\be
\mu_q=x \, \mu+\sqrt{(1-x^2)(1-\mu^2)}\cos(\phi_q)
\ee
For an arbitrary function $\psi(\mathbf {k},\mathbf{ q})$, we can write:
\be
\int \psi(\mathbf{ k},\mathbf {q})d^3q=\int_{0}^{\infty} q^2 dq\int_{-1}^{1}dx\int_{0}^{2\pi}d\phi_q \psi(\bk,\bq) \, .
\ee
Therefore the integration corresponding to $P_{22}^s$ in Eq.~(\ref{e:heavappend}) reduces to
\be
P_{22}^s = \frac{2 k^3}{(2 \pi)^3} \int dr r^2 P_{11}(r) \int_{-1}^{1} dx  P_{11}(k \sqrt{1+r^2-2r\, x}) \int_{0}^{2\pi} d\phi_q [F_2^S(r,\phi_q,x,\mu,b_1,f)]^2 \, ,
\ee
where we take the integral over $\phi_q$ analytically, and the result can be found in our Mathematica notebook.
We write $P_{22}^s$ as a sum over powers of $\mu$, $b_1$ and $f$ as
\be 
P_{22}^s = \sum_{\ell=0}^{\ell=4} \sum_{m=0}^{m=2}  \sum_{n=0}^{n=4}   \mu^{2\ell} b_1^m f^n A_{\ell m n}(r,x) \, , 
\ee
and finally by integrating over $r$ and $x$ we find the coefficients $A_{\ell m n}(r,x)$ numerically.\\
Next, we explain the computation of $P_{13}^s$, where the integration is as following:
\be
P_{13}^s = 6(1+\beta\mu^2)b_1 P_{11}(k)\frac{1}{(2\pi)^3}\int dr r^2 P_{11}(r) \int_{-1}^1 d\mu_q\int_{0}^{2\pi} d\phi_q F_3^S(q,\phi_q,x,\mu,b_1,f) \, ,
\ee

We integrate over $\phi_q$ and $\mu_q$ analytically, the result of which is contained in our Mathematica notebook.
Then similar to $P_{22}^s$, we write $P_{13}^s$ as a sum over powers of $\mu$, $b_1$ and $f$ as 
\be 
P_{13}^s = \sum_{\ell=0}^{\ell=2} \sum_{m=0}^{m=1}  \sum_{n=0}^{n=3}   \mu^{2\ell} b_1^m f^n B_{\ell m n}(r) \, , 
\ee
and by integrating over $r$ numerically, we find the coefficients $B_{\ell m n}(r)$ which are given in our Mathematica notebook.
We use the minimum and maximum values of wave number that we have from our {\sc CLASS} output, for the limits of $q$.
\par In order to avoid numerical problems, we use series expansion for large and small values of $q$.

\section{TNS model \textit{A}, \textit{B} and \textit{C} correction terms} \label{Appendix:ABC}
In this appendix we present the basic forms of the RSD correction terms appearing in Eq.~(\ref{eq:pTNS}). These terms are given as 

\begin{align}
A(k,\mu) =& \ \sum_{m,n=1}^3 \mu^{2m} f^{n} \frac{ k^3}{(2\pi)^2} \nonumber \\ 
    &\times \bigg[ \int dr \int dx \Big( A_{mn}(r,x) P_{\rm lin}(k) + \tilde{A}_{mn}(r,x) P_{\rm lin}(kr,z)\Big)  \nonumber \\ 
    &\times \frac{P_{\rm lin}(k\sqrt{1+r^2-2rx},z)}{(1+r^2-2rx)} + P_{\rm lin}(k,z)\int dr a_{mn}(r)P_{\rm lin}(kr,z) \bigg], \\
B(k,\mu) =& \ \sum_{n=1}^4 \sum_{a,b=1}^2 \mu^{2n}(-f)^{a+b}\frac{k^3}{(2\pi)^2} \nonumber \\  
    &\times \int dr \int dx     B^n_{ab}(r,x)\frac{P_{a2}(k\sqrt{1+r^2-2rx},z) P_{b2}(kr,z)}{(1+r^2-2rx)^a}, \\
C(k,\mu) =& \  (k\mu f)^2 \nonumber \\ 
    &\times \int \frac{d^3p d^3q}{(2\pi)^3}\delta_D(\bfk-\bfq-\bfp)\frac{\mu_p^2}{p^2}(1+fx^2)^2P_{\rm lin}(p,z)P_{\rm lin}(q,z),
\label{cterm}
\end{align}
where $\mu_p =\hat{\bfk}\cdot \hat{\bfp}$, $r=k/q$ and $x=\hat{\bfk}\cdot \hat{\bfq}$. Explicit expressions for $A_{mn},\tilde{A}_{mn},a_{mn}$ and $B_{ab}^n$ can be found in the Appendices of \cite{Taruya:2010mx}. The $C(k,\mu)$ term is known to have small oscillatory features and thus it is usually omitted in the literature. We choose to include it in our work.

\section{Fitting Procedure for EFT and TNS model} \label{Appendix:fit}
To fit the RSD free parameters of the EFT (Eq.~\ref{eq:pEFT}) and TNS (Eq.~\ref{eq:pTNS}) models to the simulation data we simply minimize the $\chi_{\rm red}^2$
\begin{eqnarray}
\chi^2_{\rm red}(k_{\rm max}) & = \frac{1}{N_{\rm dof}}\sum\limits_{k=k_{\rm min}}^{k_{\rm max}} \sum\limits_{\ell,\ell'=0,2} \left[P^{S}_{\ell,{\rm data}}(k)-P^{S}_{\ell,{\rm model}}(k)\right]\nonumber \\ & \times \mbox{Cov}^{-1}_{\ell,\ell'}(k)\left[P^{S}_{\ell',{\rm data}}(k)-P^{S}_{\ell',{\rm model}}(k)\right],
\label{covarianceeqn}
\end{eqnarray}
where $\mbox{Cov}_{\ell,\ell'}$ is the Gaussian covariance matrix between the different multipoles, and $k_{\rm min}=0.006 \, h/{\rm Mpc}$. The number of degrees of freedom $N_{\rm dof}$ is given by $N_{\rm dof} = 2\times N_{\rm bins} - N_{\rm params}$, where $N_{\rm bins}$ is the number of $k-$bins used in the summation and $N_{\rm params}$ is the number of free parameters in the theoretical model. Here, $N_{\rm params} = 2$ for EFT and not 3 because we only fit the first two multipoles\footnote{The inclusion of the hexadecapole would  restrict the determined range we can safely fit to. Further, the monopole and quadrupole contain most of the RSD information so we can omit the hexadecapole from these fits.}, and $N_{\rm params} = 1$ for the TNS model. 

We increase $k_{\rm max}$ until $\chi^2_{\rm red}(k_{\rm max}) \geq 1$. This gives a good indication of where the model doesn't fit the data so well anymore. In the fit we keep cosmology fixed to the COLA simulation's fiducial values and so only vary the counter term coefficients and $\sigma_v$. 

We apply linear theory to model the covariance between the multipoles (see Appendix C of \cite{Taruya:2010mx} for details). This has been shown to reproduce N-body results up to $k\leq 0.300h/\mbox{Mpc}$ at $z=1$. In the covariance matrix we assume a number density of $n= 1\times 10^{-3} \, h^3/\mbox{Mpc}^3$ and a survey volume of $V_s=4 \, \mbox{Gpc}^3/h^3$ which are similar specifications for a Euclid like survey \cite{Amendola:2016saw}. The best fit parameters as well as $k_{\rm max}$ are shown in Table~\ref{fittable}.

\begin{table}
\centering
\caption{Table showing the maximum $k_{\rm max} [h/{\rm Mpc}]$ used in Eq.~(\ref{covarianceeqn}) and best fit model parameters for TNS and EFT models found by a least $\chi^2$ fit to the COLA data.  \vspace{0.2cm}}
\begin{tabular}{| c || c | c | c | c || c | c | c | c |}
\hline  
 \multicolumn{1}{ | c || }{Model} & \multicolumn{4}{|c||}{TNS} &  \multicolumn{4}{|c|}{EFT} \\
 \hline
 z & $0.5$ & $ 1 $ & $1.5$ & $2$ & $0.5$ & $1$ &  $1.5$ & $2$ \\ \hline\hline 

 $k_{\rm max}$ & $0.16$ & $0.21$ & $0.27$ & $0.35$  & $0.16$ & $0.21$ & $0.27$ & $0.311$  \\ \hline 
 $\sigma_v$ & $7.35$ & $6.26$ & $5.12$ & $4.19$  & - & - & - & -   \\ \hline 
 ${c_{2|\delta_s,0}}/{k_{\rm nl}^2}$ & - & - & - & -  & $0.05$ & $0.00$  & $0.00$ & $0.13$ \\ \hline 
 ${c_{2|\delta_s,2}}/{k_{\rm nl}^2}$ & - & - & - & -  & $13.57$ & $8.96$  & $5.66$ & $1.52$ \\ \hline 
 ${c_{2|\delta_s,4}}/{k_{\rm nl}^2}$ & - & - & - & -  & $7.34$ & $8.03$ & $6.86$ & $5.73$ \\  
\hline
\end{tabular} \vspace{0.2cm}
\label{fittable}
\end{table}

\section{Neglecting the lensing term\label{app:lensing}}
Throughout the paper, we have neglected the lensing contribution to the angular power spectrum. In this appendix, we show that among the three different redshift bins that we used, namely, $\Delta z = 0.1$, $\Delta z = 0.01$, and $\Delta z = 0.001$, lensing is of the same order as the RSD contribution  for $\Delta z = 0.1$ while for the other two redshift bins, it is negligible. 
\begin{figure}[ht]
    \centering
    \includegraphics[scale=0.5]{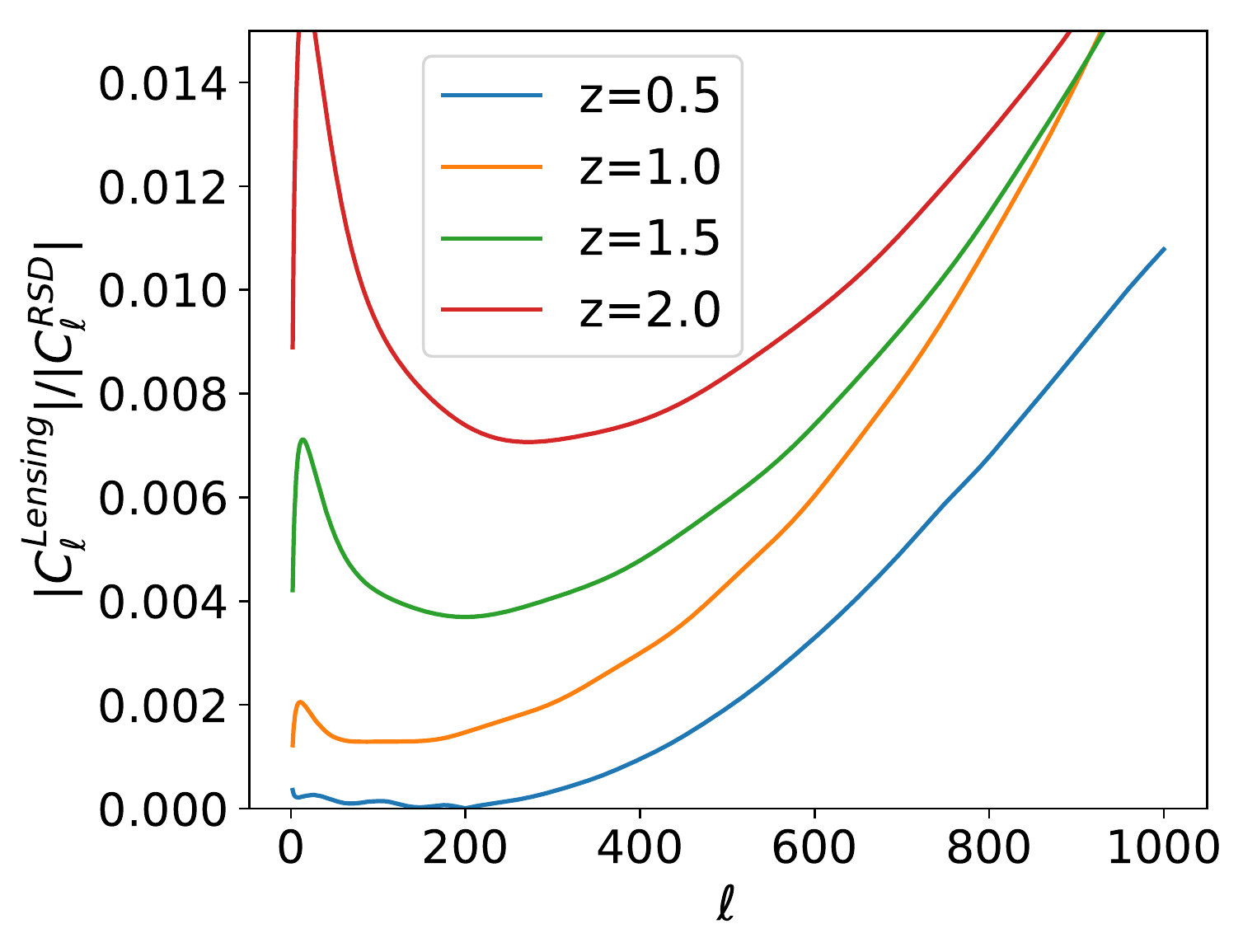}
    \includegraphics[scale=0.5]{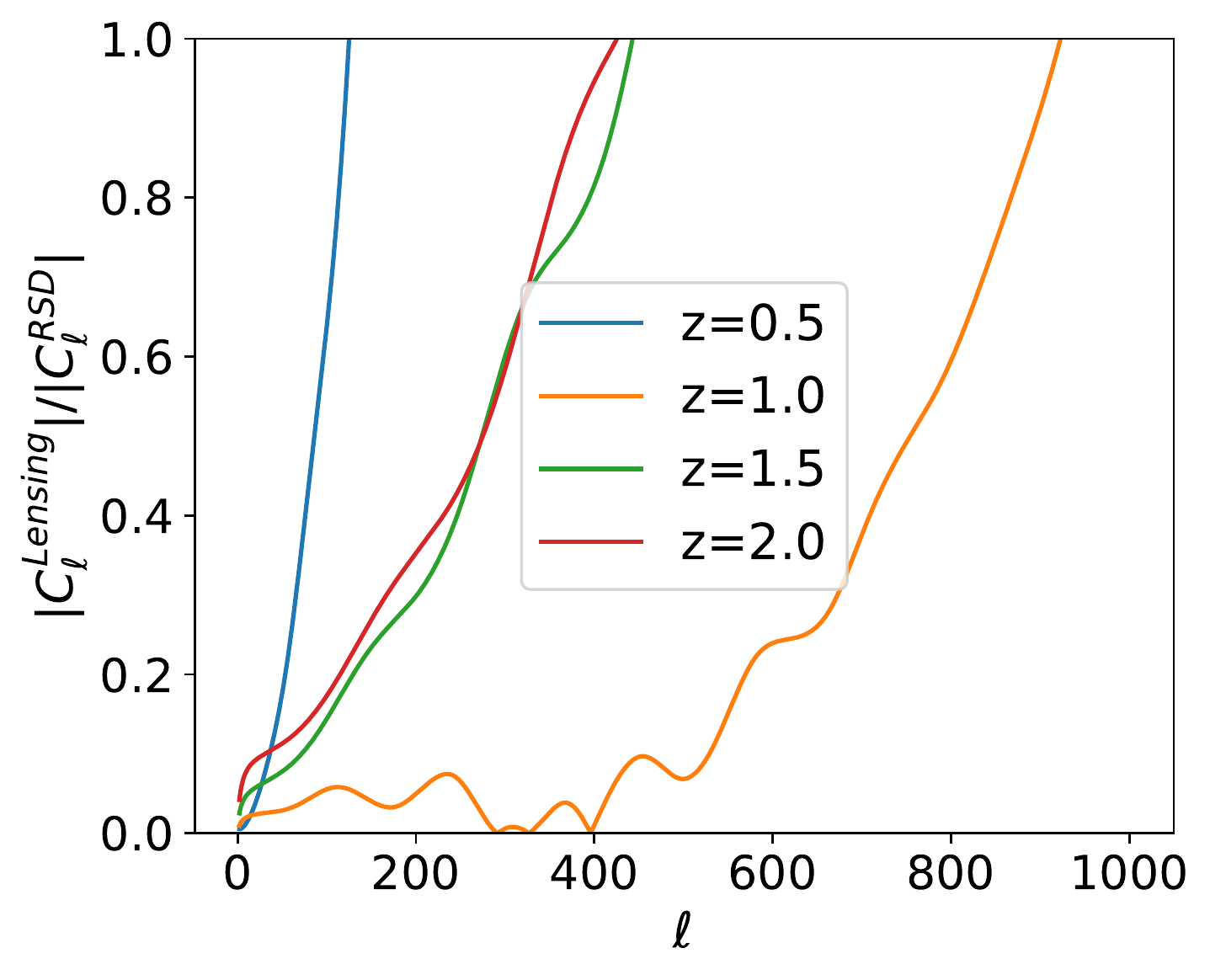}
    \caption{Ratio of lensing terms to RSD tems for different redshifts with $\Delta z = 0.01$ (left panel) and $\Delta z = 0.1$ (right panel). }
    \label{fig:RSD_vs_lensing}
\end{figure}

In Fig.\ \ref{fig:RSD_vs_lensing}, we show the ratio of lensing to the RSD term for different redshifts with $\Delta z = 0.01$ (left panel) and $\Delta z = 0.1$ (right panel). For $\Delta z = 0.1$, lensing is not negligible which compared to RSD, however, we have shown that for this window width, RSD effect is also not very significant. For $\Delta z  = 0.001$ (left panel), we can see that lensing terms are at most $1\%$ of RSD terms.
It is also interesting to note that for $\De z=0.1$ the lensing signal is very small at  $\ell<400$. This comes from the fact that the lensing signal is the sum of the always negative  lensing-density correlation and the positive lensing-lensing term. As the density term is larger than lensing, at low redshift the signal is dominated by the first term and is therefore negative. At sufficiently high redshift when enough lensing has accumulated, the lensing-lensing term starts to dominate and the signal becomes positive. For $\De z=0.1$ this happens roughly at $z=1$. For $\De z=0.01$ this happens roughly at $z\sim 0.5$ for the low multipoles, $\ell<200$ while for higher multipoles the positive lensing-lensing signal dominates. Since the cross correlation lensing$(z_2)$-density$(z_1)$ is significant only for density fluctuations at redshift over which the lensing term is integrated, $z_1<z_2$, this contribution is smaller for smaller redshift bins.
\section{Fisher forecast\label{app:Fisher}}
In this appendix, we explain in more details the Fisher forecast we have done for RSD detection. We replace each $\mu^2$ term with $A_\mu \mu^2$, where $A_\mu$ is an artificial amplitude with fiducial value of 1, and our aim is to forecast how precisely we can measure this amplitude. For non-linear RSD, we simply use the Kaiser formula applied to halofit model, basically replacing $P(k_{\parallel}, \frac{\ell}{\chi}, \bar z)$ in Eq.~(\ref{eq:flat-sky-Ruth}) with the Kaiser formula given in Eq.~(\ref{e:poldk}), we have
\begin{align}
C_\ell^{\Delta \Delta}( z,z) & = \frac{1}{2 \pi \chi^2(\bar{z})} \left[ \int_{-\infty}^{+\infty} d k_{\parallel} (1+  2 A_{\mu} f \mu^2 + A_{\mu}^2 f^2 \mu^4)P(k_{\parallel}, {\ell}/{\chi} , z ) \right] \nonumber \\
& \equiv C_\ell^{\delta \delta} + A_{\mu} C_\ell^{\delta \theta} + A_{\mu}^2 C_\ell ^{\theta \theta} \, ,
\end{align}
where $\Delta$ is the density perturbations in redshift space
\be \Delta(\bn , z) = \delta(\bn , \mathbf{r}) - \frac{\nabla_z v_z(\mathbf{r})}{a H(z)} \, ,\ee
and $\theta = \nabla_z v_z(\mathbf{r})/a H(z) $, where $z$ is the line of sight direction.
For the Fisher forecast, we follow a similar approach as the one used in Section 4 of \cite{Jalivand:2018vfz}. The Fisher matrix for parameters $\alpha$ and $\beta$ with covariance matrix, $\mathcal{C}$, follows the formula
\be
F_{\alpha \beta}=\sum_{\ell} \frac{2 \ell+1}{2} \left[\left(\partial_{\alpha} \mathcal{C}\right)\left(\mathcal{C}^{-1}\right)\left(\partial_{\beta} \mathcal{C}\right)\left(\mathcal{C}^{-1}\right)\right]
\ee
which for our case with one parameter, $A_{\mu}$ and covariance matrix being the $C_\ell$'s,  reduces to 
\begin{align}
F_{A_{\mu} A_{\mu}}=\sum_{\ell=2}^{\ell_{\rm max}} \frac{2 \ell+1}{2} \left[\frac{\partial_{A_{\mu}} C_{\ell}^{\Delta\Delta} } {C_{\ell}^{\Delta\Delta} }\right]^2 \, .
\end{align}
\bibliographystyle{JHEP}
\bibliography{biblio}
\end{document}